\documentclass{aa}
\usepackage{graphicx}
\usepackage{natbib}
\bibpunct{(}{)}{;}{a}{}{,} %
\normalfont
\newcommand{\kms}{\rm ~km~s^{-1}}

\newcommand{\cc}{\mbox{ cm}^{-3}}

\newcommand{\ang}{\mbox{ \AA}}
\newcommand{\etal}{\rm et al.}
\def\EE#1{\times 10^{#1}}

\def\ccm{\rm ~cm^{-3}}
\def\cm2{\rm ~cm^{-2}}
\def\kms{\rm ~km~s^{-1}}

\def\ergsm{\rm ~erg~s^{-1} cm^{-2}}
\def\ergsma{\rm ~erg~s^{-1} cm^{-2} \AA^{-1}}
\def\wl{~\lambda}
\def\wll{~\lambda\lambda}

\def\KK{\rm ~K}
\def\Ti44{M(^{44}{\rm Ti})}

\def\lsim{\!\!\!\phantom{\le}\smash{\buildrel{}\over
  {\lower2.5dd\hbox{$\buildrel{\lower2dd\hbox{$\displaystyle<$}}\over
                               \sim$}}}\,\,}
\def\gsim{\!\!\!\phantom{\ge}\smash{\buildrel{}\over
  {\lower2.5dd\hbox{$\buildrel{\lower2dd\hbox{$\displaystyle>$}}\over
                               \sim$}}}\,\,}

\begin{document}
   \title{Time evolution of the line emission from the 
   inner circumstellar ring of SN 1987A and its hot spots
   \thanks{Based on observations made with ESO telescopes at the La Silla
     Paranal Observatory under programme IDs 60.A-9022, 66.D-0589, 70.D-0379,
     074.D-0761, 078.D-0521, 080.D-0727.}}

%   \subtitle{}

 \author{Per Gr\"oningsson\inst{1}
     	\and
   	Claes Fransson\inst{1}
	\and  
	Bruno Leibundgut\inst{2}
	\and 
        Peter Lundqvist\inst{1}
	\and  
        Peter Challis\inst{3}
        \and
        Roger A. Chevalier\inst{4}
	\and  
 	Jason Spyromilio\inst{2}
    	}

   \offprints{P. Gr\"oningsson}

   \institute{
	Stockholm Observatory, Stockholm University,
              AlbaNova University Center, SE-106 91 Stockholm, Sweden\\
              \email{per@astro.su.se, claes@astro.su.se, peter@astro.su.se}
	\and
	European Southern Observatory,
	Karl-Schwarzschild-Strasse 2, D-85748 Garching, Germany
	\and 
        Harvard-Smithsonian Center for Astrophysics, 60 Garden Street, MS-19,
        Cambridge, MA 02138.
        \and
        Department of Astronomy, University of
        Virginia, P.O. Box 400325, Charlottesville, VA 22904, U.S.A.
	}

   \date{Received; accepted}

\titlerunning{Line emission from the inner ring of SN~1987A}
\authorrunning{Gr\"oningsson et al.}

   \abstract{
We present seven epochs between October 1999 and November 2007 of high
resolution VLT/UVES echelle spectra of the ejecta-ring collision of SN 1987A.

The fluxes of most of the narrow lines from the unshocked gas decreased by a
factor of $2-3$ during
this period, consistent with the decay from the initial ionization by the shock
break-out. However, [O~III] in particular shows an increase up to day
$\sim6800$. This agrees with radiative shock models where the
pre-shocked gas is heated by the soft X-rays from the shock.
The evolution of the [O~III] line ratio shows a decreasing temperature of the
unshocked ring gas, consistent with a transition from a hot, low density
component which was heated by the initial flash ionization to the lower
temperature in the pre-ionized gas ahead of the shocks.

The line emission from the shocked gas increases rapidly as the shock sweeps up
more gas. We find that the neutral and high ionization lines follow the
evolution of the Balmer lines roughly, while the intermediate ionization lines
evolve less rapidly. Up to day $\sim6800$, the optical light curves have a
similar evolution to that of the soft X-rays. The break
between day $6500$ and day $7000$
for [O~III] and [Ne~III] is likely due to recombination to lower ionization
levels. Nevertheless, the evolution of the [Fe~XIV] line, as well as the lines
from the lowest ionization stages, continue to follow that of the soft X-rays,
as expected.

There is a clear difference in the line profiles between the low and
intermediate ionization lines, and those from the coronal lines at
the earlier epochs. This shows that these lines arise from regions with
different physical conditions, with at least a fraction of the coronal lines
coming from adiabatic shocks.
At later epochs the line widths of the low ionization lines, however, increase
and approach those of the high ionization lines of [Fe X-XIV].
The H$\alpha$ line profile can be traced up to $\sim500\kms$ at the latest
epoch. This is consistent with the cooling time of shocks propagating into
a density of $(1-4)\EE4\cc$. This means that these shocks are among the highest
velocity radiative shocks observed.

   \keywords{supernovae: individual: SN 1987A --
                circumstellar matter, shocks
               }
   }

   \maketitle
	
%
%________________________________________________________________

\section{Introduction}
Since the explosion more than two decades ago, supernova (SN) 1987A in
the Large Magellanic Cloud (LMC) has been extensively observed in
multiple wavelength ranges.
It is now by far the most studied of all supernovae.
A few months after outburst, the first evidence of the stationary
circumstellar matter around the SN came from detection of narrow
optical lines \citep{wampler88} and
UV emission lines by the International Ultraviolet Explorer (IUE)
\citep{fran89}. Resolved imaging observations later disclosed
the now classical ring nebula around the SN consisting of three
approximately plane-parallel rings; the radioactive SN debris is
located at the center of the inner equatorial ring (ER). The ER is
intrinsically close to circular in shape and has a tilt angle of
$43^\circ$ \citep{sugerman05,kjaer07}. The two outer rings (ORs) are displaced
by $\sim1.3~{\rm ly}$ \citep{sugerman05} on either side of the ER ring
plane \citep{crotts89,wampler90,burrows95}. The observed radii
of the rings ($\sim0.6~{\rm ly}$ for the ER), together with the
measured radial expansion velocities ($\sim10.3\kms$ for the ER),
indicate that the ring system was ejected about $20,000$ years ago, if
constant expansion velocities of the rings are assumed.
The formation mechanism of the three rings is still not fully understood,
although
a scenario is that the structure was formed as a result of a
merger with a binary companion in a common envelope phase \citep{MP07}.
However, single star models for the progenitor have also been proposed
\citep[e.g.,][]{eriguchi92,woosley97}.

The ring structure was photoionized by the EUV and soft X-ray flash
emitted from the SN explosion, and since then the gas has been cooling,
recombining and fading roughly linearly with time
\citep[e.g.,][]{LF96}. This gas can be traced by narrow emission lines
and the fading of these lines is consistent with ring gas
densities in the range $3\EE3-3\EE4\ccm$ \citep{LF96,pun07}. Early light
echoes revealed, however, that the observed photoionized gas is probably only a
small fraction of the total mass ejected from the progenitor system
\citep{gouiffes88,crotts89}.\\

A new phase was entered, marking the birth of the SN remnant, as
the SN debris drives a blast wave into its circumstellar medium.
The first indication of the encounter between this blast wave and
the ER came in 1997 from HST/STIS observations where broad ($\sim 250\kms$)
blue-shifted emission components were detected in H$\alpha$
\citep{sonneborn98}.
Later it was found that this interaction could be
detected in HST images already in 1995 \citep{law00}. The interaction
took place in a small region located in the northeastern part of the
ER (P.A. $29^\circ$) \citep{pun97}. This ``hot spot'' is usually
referred to as Spot 1.  Hence, Spot 1 is believed to mark the place
were the blast wave first strikes an inward-pointing protrusion of the ER
surrounded by an H II region inside the ring \citep{pun02}. The
protrusions are presumably a result of Rayleigh-Taylor instabilities
caused by the interaction of the stellar progenitor wind with the ER. Since
1997, Spot 1 has steadily increased in flux and many more clumps have been
shocked giving rise to more ``hot spots'' in
different locations around the ring. Ten years later, there are ``hot
spots'' distributed all around the ER.
It is, however, not clear if the ring is really
homogeneous, or if it may consist of isolated clumps. Time will
tell.\\

In the interaction between the supernova ejecta and the circumstellar
matter a multi-shock system develops. A forward shock (blast wave)
propagates into the CSM and a reverse shock is driven backward (in the
Lagrangian frame of reference) into the ejecta \citep{chevalier82}. A
complication is that the circumstellar structure is probably complex, with a
constant density H II region inside the ER \citep{CD95}. When
the forward shock, propagating with a velocity of $\sim3500\kms$
through the circumstellar gas (with density $\sim10^2\ccm$), reaches
the denser ER ($\sim10^4{\rm ~cm^{-3}}$), slower shocks are transmitted into
the ring. Because the transmitted shock velocities depend on both the
density profile of the ER and the incident angle, a wide range
of shock velocities is expected ($\sim10^2-10^3\kms$) \citep{michael02}. In
addition,  the interaction with a density jump 
may create reflected shocks between the ring and the reverse shock.\\

The interaction has been observed in virtually all wavelength ranges, from
radio to X-rays \citep[e.g.][]{mccray07}.
The high resolution X-ray spectrum observed by Chandra shows a number
of H and He-like emission lines in the wavelength range $5-25$~\AA~such as
Si XIII-XIV, Mg XI-XII, Ne IX-X, O VII-VIII, Fe XVII and N VII emitted by
shocks produced by the interaction of the blast wave and the ER. The line
profiles indicate gas velocities in the range $\sim300-1700\kms$
\citep{zhekov05,zhekov06,dewey08}.
As the expanding ejecta collide with the CSM, the luminosity increases rapidly
in radio synchrotron emission \citep{gaensler07}, as well as in
X-rays originating from the hot shocked gas \citep{park07}. 
The ring evolution has also been studied in mid-IR \citep{bouchet06},
near-IR \citep{kjaer07}, as well as optical and UV \citep{pun02,sugerman02,
sugerman05,Gro06,Gro08}.\\

For this paper, the X-rays are of special interest.
Up to 1999 both the soft and hard X-rays correlated
well with the radio emission. Since then, however, the flux of soft
X-rays began to increase much more rapidly than both radio and hard
X-ray fluxes \citep{park07}. Instead, in \citet{Gro06} we found that the
evolution of the flux of the
optical high ionization lines from the hot spots correlated
well with the soft X-ray emission.  This suggests that much of the
soft X-ray emission arises from the ejecta-ring collision, while the hard
X-rays mainly come from the hot gas between the reverse and the
forward shock. The radio emission continues to correlate well with
hard X-rays, and is most likely caused by relativistic electrons
accelerated by the reverse shock interior to the ER \citep{zhekov06}.

A complementary view of the interaction is offered by the IR emission
from the dust \citep[][]{bouchet06,dwek08}. Dwek et al. find a temperature
of the silicate rich dust of $\sim 180$ K, heated either by the
shocked gas or by the radiation from the shocks. The dust emission
coincides roughly with the optical emission. A surprising result is
the comparatively low IR to X-ray ratio, indicating destruction of the
dust by the shock.  The fact that this ratio decreased with time is
also consistent with dust destruction.

In this paper we discuss the evolution of the optical emission lines from high
resolution spectroscopic data taken with VLT/UVES. In the first paper in
this series \citep{Gro06} we discussed the implications of the coronal
lines from [Fe X-XIV] found in these spectra. In the second paper
\citep[][ in the following referred to as G08]{Gro08} we discussed a full
analysis of the spectrum from
one epoch, 2002 October. These spectra showed narrow emission line
components with FWHM $\sim 10-30\kms$ for most of the lines, arising
from the unshocked ring gas. The interaction between the ejecta and
the ER can be seen as intermediate velocity lines coming from shocked ring
material, extending to
approximately $\pm300\kms$, with clearly asymmetric line profiles. In addition,
a broad ($\sim15,000\kms$) velocity component in H$\alpha$ and [Ca~II],
originating from the reverse shock in the outer ejecta \citep{heng07}, is
clearly visible. In
contrast to the analysis in G08, we focus in this paper on the time evolution
of the line emission. Hence, in this paper, we present data taken with VLT at
seven epochs ranging from 1999 October to 2007 November.\\

The observations and data reductions are discussed in Sect.
\ref{sec:observations}.
Analysis of the data is presented in Sect. \ref{sec:analysis} followed by
discussions of the results in Sect. \ref{sec:discussion}, and finally a summary
in Sect. \ref{sec:summary}.

%__________________________________________________________________

\section{Observations}
\label{sec:observations}

High resolution spectra of SN 1987A were obtained using the Ultraviolet
and Visual Echelle Spectrograph (UVES) at ESO/VLT at Paranal, Chile
\citep{dekker00}. The UVES
spectrograph disperses the light beam in two separate arms covering different
wavelength ranges of the spectrum. The blue arm covering the shorter
wavelengths ($\sim3030-4990$ \AA) is equipped with a single CCD detector
with a spatial resolution of $0\farcs246/\mathrm{pixel}$ while
the red arm ($\sim4760-10\,600$ \AA) is covered by a mosaic of two CCDs having
a resolution of $0\farcs182/\mathrm{pixel}$. Thus with two different
dichroic settings the wavelength coverage is $\sim3030-10\,600$ \AA. Due to
the CCD mosaic in the red arm there are, however, gaps at $5770-5830$~\AA~and
$8540-8650$~\AA. The spectral resolving power is $40\,000$ for a $1\farcs0$
wide slit.\\

\begin{figure*}
\resizebox{\hsize}{!}{\includegraphics{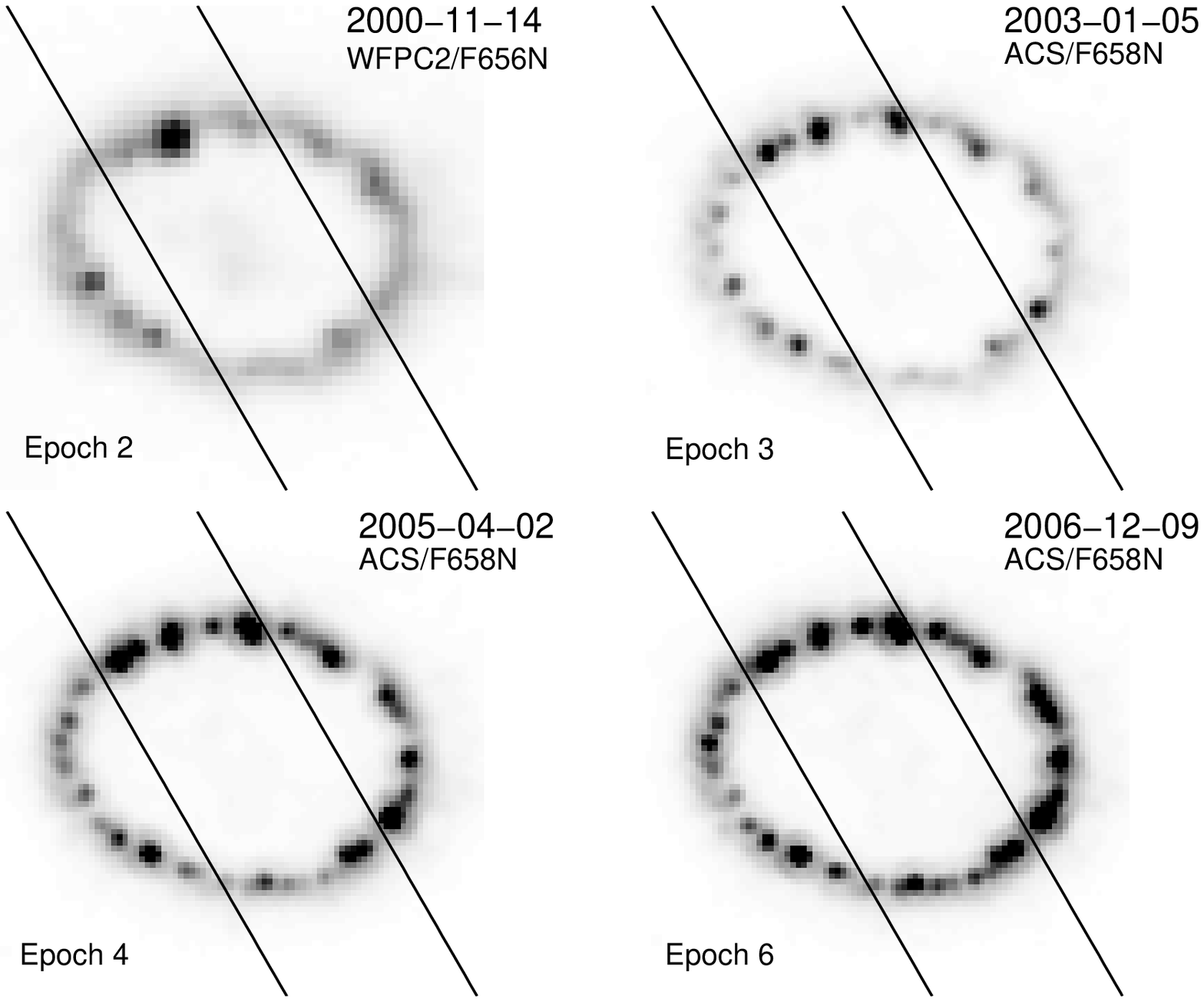}}
\caption{Ring images taken with HST at four different epochs
(obtained by the SAINTS team; PI: R.P. Kirshner).
The slit position of the VLT/UVES observations is superimposed (the slit width
is $0\farcs8$ and $\rm{PA}=30^\circ$).
North is up and east is to the left. All images except the earliest epoch
(upper left panel) have the same scaling in flux. The epochs labeled in the
panels correspond roughly to the time of the HST observations. Note the
evolution and distribution of ``hot spots'' around the ring. The evolution is
most prominent on the southwest part of the ring.}
\label{fig:ringimage_hst}
\end{figure*}

Spectra of SN 1987A were obtained at 7
different epochs ranging from October 16 1999 to November 28 2007,
henceforth referred to as Epochs 1--7. See Table 1 for the
observational details. The observations were performed in service mode
for Epochs 2--7 whereas Epoch 1 was part of the instrument ``first
light'' program. For Epochs 2--7 a $0\farcs8$ wide slit was
centered on the SN and with a position angle PA$=30^{\circ}$ (see Fig.
\ref{fig:ringimage_hst}). Hence,
the spectral resolution for these epochs was
$\lambda/\Delta\lambda\sim50\,000$ which corresponds to a velocity of
$6\kms$. For Epoch 1 the slit width was $1\farcs0$ and rotated to
PA$=20^{\circ}$.

\begin{table*}
\centering
\caption{VLT/UVES observations of SN 1987A and its rings.}
\begin{tabular}{l l c c c c c c c c}
\hline
\hline
Epoch&Date&Days after&Setting&$\lambda$ range&Slit
width&Resolution&Exposure&Seeing&Airmass\\ 
&&explosion&&(nm)&(arcsec)&$(\lambda/\Delta\lambda$)&(s)&(arcsec)\\
\hline
1&1999 Oct 16&4618&346+580&303--388&1.0&40,000&1,200&1.0&1.4\\
&&&&476--684&&&&\\
&&&&&&&&\\
2&2000 Dec 9--14&5039--5043&346+580&303--388&0.8&50,000&10,200&0.4--0.8&1.4\\
&&&&476--684&&&&\\
&&5038--5039&390+860&326--445&&&9,360&0.4&1.4--1.6\\
&&&&660--1060&&&&\\
&&&&&&&&\\
3&2002 Oct 4--7&5704--5705&346+580&303--388&0.8&50,000&10,200&0.7--1.0&1.5--1.6\\
&&&&476--684&&&&\\
&&5702--5703&437+860&373--499&&&9,360&0.4--1.1&1.4--1.5\\
&&&&660--1060&&&&\\
&&&&&&&&\\
4&2005 Mar 21&6601--6623&346+580&303--388&0.8&50,000&9,200&0.6--0.9&1.5--1.8\\
&Apr 8-12&&&476--684&&&&\\
&&6621--6623&437+860&373--499&&&4,600&0.5&1.6--1.7\\
&&&&660--1060&&&&\\
&&&&&&&&\\
5&2005 Oct 20&6826&346+580&303--388&0.8&50,000&2,300&0.9&1.4\\
&Nov 1--12&&&476--684&&&\\
&&6814--6837&437+860&373--499&&&9,200&0.5--1.0&1.4--1.6\\
&&&&660--1060&&&&\\
&&&&&&&&\\
6&2006 Oct 1--29&7160--7180&346+580&303--388&0.8&50,000&9,000&0.5--0.9&1.4--1.5\\
&Nov 10--15&&&476--684&&&\\
&&7188--7205&437+860&373--499&&&9,000&0.5--1.0&1.4\\
&&&&660--1060&&&&\\
&&&&&&&&\\
7&2007 Oct 23--24&7547--7583&346+580&303--388&0.8&50,000&11,250&0.8--1.4&1.4--1.6\\
&Nov 27--28&&&476--684&&&\\
&&7547&437+860&373--499&&&9,000&1.1--1.4&1.4--1.6\\
&&&&660--1060&&&&\\

\hline
\hline
\label{tab:obslog}
\end{tabular}
\end{table*}

For the data reduction we made use of the MIDAS implementation of the UVES
pipeline version 2.0 for Epochs 1--3 and version 2.2 for Epochs 4--7.
For a detailed description of the steps involved in the reductions we refer to
G08.

For the time evolution of the fluxes an accurate absolute flux
calibration is necessary. As discussed in \citet{Gro06} and G08, we estimated the accuracy
of the flux calibration by comparing the spectra of flux-calibrated
spectrophotometric standard stars with their tabulated physical fluxes. In
addition, we compared emission line fluxes from different exposures (and hence
with different atmospheric seeing conditions) within the same epochs. From
these measurements, together with the results from HST photometry in
Appendix \ref{sec:flux_comp}, we conclude that the accuracy in relative
fluxes should be within $10-15\%$.
The estimation of the absolute systematic flux error was done from a comparison
between the UVES fluxes and HST spectra and photometry (see G08 and Appendix
\ref{sec:flux_comp}). From this,
together with our experience with such data,
we estimate that the uncertainty of the absolute
fluxes should be less than $20-30\%$.\\

We sampled the ring at the northeast and southwest positions
(see Fig. \ref{fig:ringimage_hst}).
However, the limited spatial resolution of these observations makes it
impossible to distinguish between different hot spots located on the
same side of the ring and covered by the slit. Hence, we studied the
HST images taken at roughly the same epochs (see Fig. \ref{fig:ringimage_hst}
and Appendix) to identify
the number of hot spots covered by the slit for both each epoch and
side of the ring. For Epochs 1--2 we used images taken with the WFPC2
and narrow-band filter F656N. These images reveal that for Epoch 1
only Spot 1 contributes to the emission from the shocked gas over the
entire ring. At Epoch 2 there are contributions from effectively two
different spots at the northern side of the ring, even though Spot 1 clearly
dominates. On
the southern side, two spots have now appeared. For Epochs
3--6 we investigated images taken with the ACS instrument. These show
that at Epoch 3 the shocked gas emission on the northern part now comes
from four spots, while no new spots have appeared on the southern side. At
Epochs 4--6 there are hot spots almost all around the ring, although
the main contributions come from five spots on the northern side and four on
the southern side of the ring. At these three epochs spot 1 appears not
to be the strongest one on the northern side, although the first two spots
to appear on the southern side are still strongest at Epoch 6. Unfortunately
we have no comparable HST imaging for Epoch 7, and consequently no study of
individual spots can be made for that epoch.

%___________________________________________________________________

%\section{The data}

%___________________________________________________________________

\section{Analysis}
\label{sec:analysis}

\subsection{Systemic velocity and ring expansion}
The systemic velocity of SN 1987A can be estimated directly from the
peak velocities of the unshocked gas. The best accuracy should in
principle be obtained for the strongest emission lines with well
defined intensity peaks, such as the [O~II], [S~II] and [N~II]
doublets. However, since there are uncertainties in the rest
wavelengths and in the wavelength calibration (amounting to $\sim 0.01
\ang$, corresponding to $0.6 \kms$ at 5000 \AA), we have chosen to
include all lines in Tables \ref{tab:narrowlines2}--\ref{tab:narrowlines7}
in the velocity
estimate.  Taking the average of the peak velocities of the northern
($281.66\pm0.07\kms$) and southern ($291.81\pm0.07\kms$) parts of the ER
in the data set we find a center of mass velocity of $286.74\pm0.05\kms$
($1\sigma$ errors). This result is in agreement with \citet{wampler89}
and \citet{meaburn95} who obtained $286.3\pm0.8 \kms$ and
$286.5\pm1\kms$, respectively, but lower than $287.6\pm0.1\kms$
and $289.2\kms$ as reported by \citet{cumming94} and \citet{CH91},
respectively.

The expansion velocity of the ER is more difficult to measure directly 
from the data. One method to estimate the expansion velocity would be to
assume a luminosity and a temperature distribution of the ER and
then model the resulting line profiles of either side of the ER by
taking into account the instrumental broadening, spectral resolution
and atmospheric seeing \citep[e.g.,][]{meaburn95}. The best fit model to
the narrow emission components would then give the expansion velocity.
A more direct approach is to measure differences between the peak velocities
of the emission lines from the unshocked gas at the north and south parts of
the ER and correct for the ring inclination and the slit orientation.
From geometrical considerations we would ideally have for 
the radial velocity

\begin{equation}
V_r = {|\Delta V_{peak}|\over 2 {\rm sin}(i) {\rm cos}(\psi)}.
\label{eq:ringvel}
\end{equation}

\noindent The angle $\psi$ is given by
${\rm tan}(\psi) = {\rm tan}({\rm PA}+\phi) {\rm cos}(i)$
where $i$ is the inclination angle of the ring, PA the
position angle of the slit, and $\phi$ the angle between the minor
axis of the ring and north. However, due to the slit width and the
seeing, different parts of the ring contribute to the different
components along the line of sight. Since the total broadening is a
convolution of the macroscopic motion and the thermal motion, the peak
of the emission line will be velocity shifted with an amount and a
direction that depends on the skewness of the macroscopic velocity
distribution, temperature and the atomic weight of the element.  The
velocity distribution of the northern part of the ring is expected to
be positively skewed and the distribution at the southern part is
expected to be negatively skewed. Hence, in this case the measured
$\Delta V_{peak}$ would be systematically underestimated and, as a
consequence, Eq. (\ref{eq:ringvel}) only provides a lower limit
to the radial velocity. 
To account for the fact that the emission lines sample a range of
$\psi$'s, Eq. (\ref{eq:ringvel}) can be multiplied by the factor
$(\Delta\psi/2)/{\rm sin}(\Delta\psi/2)$. Comparing the width of the slit to
the size of the ER, we estimate $\Delta\psi$ to be in the range
$\sim60^\circ-90^\circ$. Hence, it follows that Eq. (\ref{eq:ringvel}) would
underestimate the real expansion velocity by $\sim5-10\%$.

To minimize the systematic errors, we chose to only
include data from Epoch 2 (2000 Dec. Table \ref{tab:narrowlines2}).
The observations at this epoch have the best seeing conditions
(Table \ref{tab:obslog}) and least contamination of emission from the shocked
ring gas. Furthermore,
to reduce the systematic uncertainties due to temperature effects,
we only included
the emission lines from elements with a large atomic weight.
The thermal broadening for these
lines is very small compared to the macroscopic velocity distribution.
Of course, we need to assume that these lines have roughly the same
spatial luminosity distributions as the other lines around the
ring. Therefore, we used the narrow components of the relatively
strong [S~II] $\lambda4069+76,\lambda6716+31$, [Ar~III] $\lambda7136$,
[Fe~II] $\lambda7155$ and [Ca~II] $\lambda7324$ lines in Table
\ref{tab:narrowlines2} to estimate $\Delta V_{peak}$.
If we adopt the geometrical values of the ER
suggested by \citet{sugerman02,sugerman05}, i.e., $i=43^\circ$ and
$\phi=9.6^\circ$, we derive for $\mathrm{PA}=30^\circ$ the radial expansion
velocity
$V_r = 10.3\pm0.3\kms$ ($1\sigma$ error). This result agree with earlier
estimates by \citet{cumming94} who found $10.3\pm0.4\kms$ and \citet{CH00}
who reported $10.5\pm0.3\kms$.

\subsection{Evolution of the unshocked components}

As discussed in detail in G08 (see Table 2 in that paper), we detect narrow
components for a large range of ionization stages, from neutral up to Ne~V
and Fe~VII. The only exceptions are the coronal lines from Fe~X-XIV.

\begin{figure}
\resizebox{\hsize}{!}
{\includegraphics[scale=0.1]{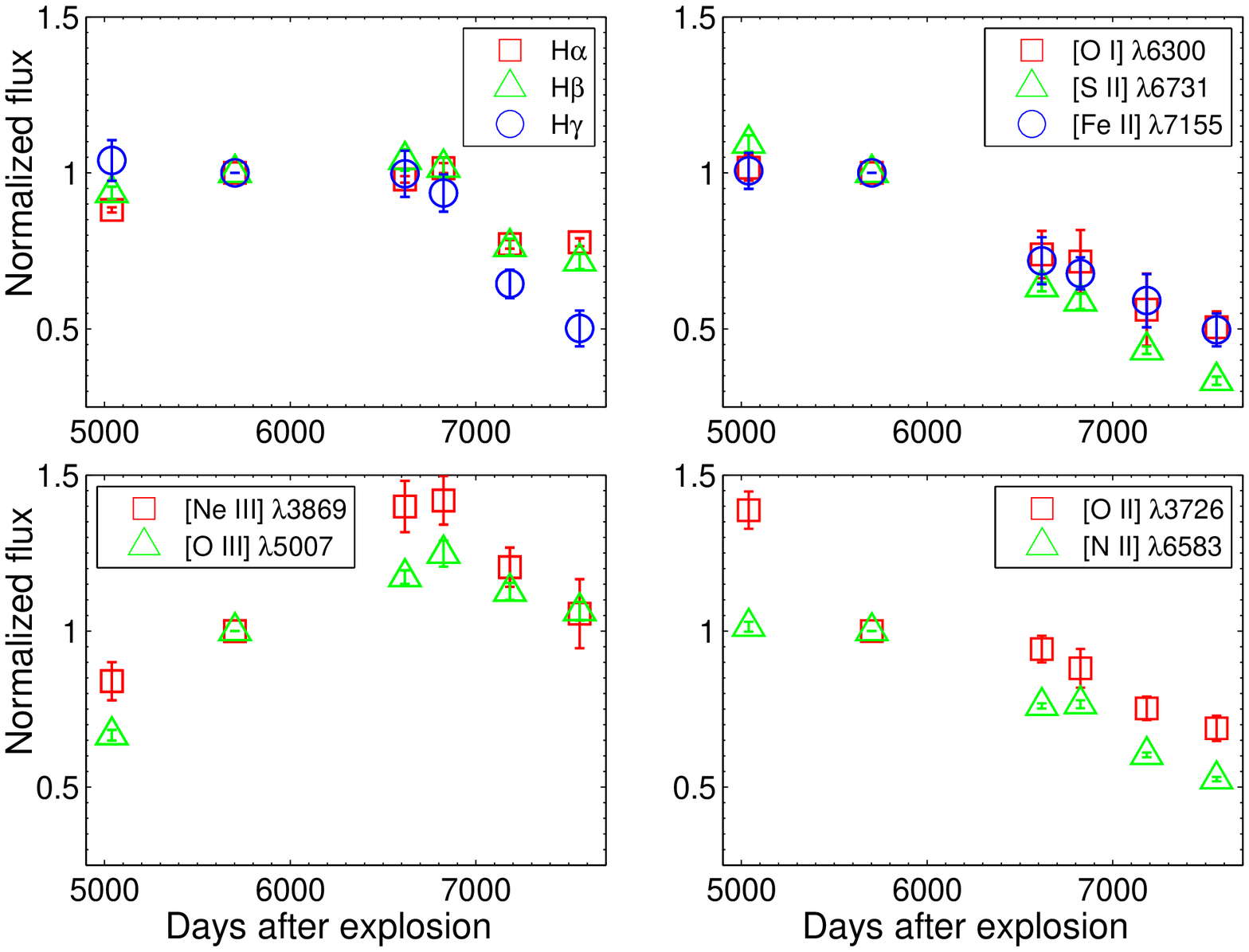}}
{\includegraphics[scale=0.445]{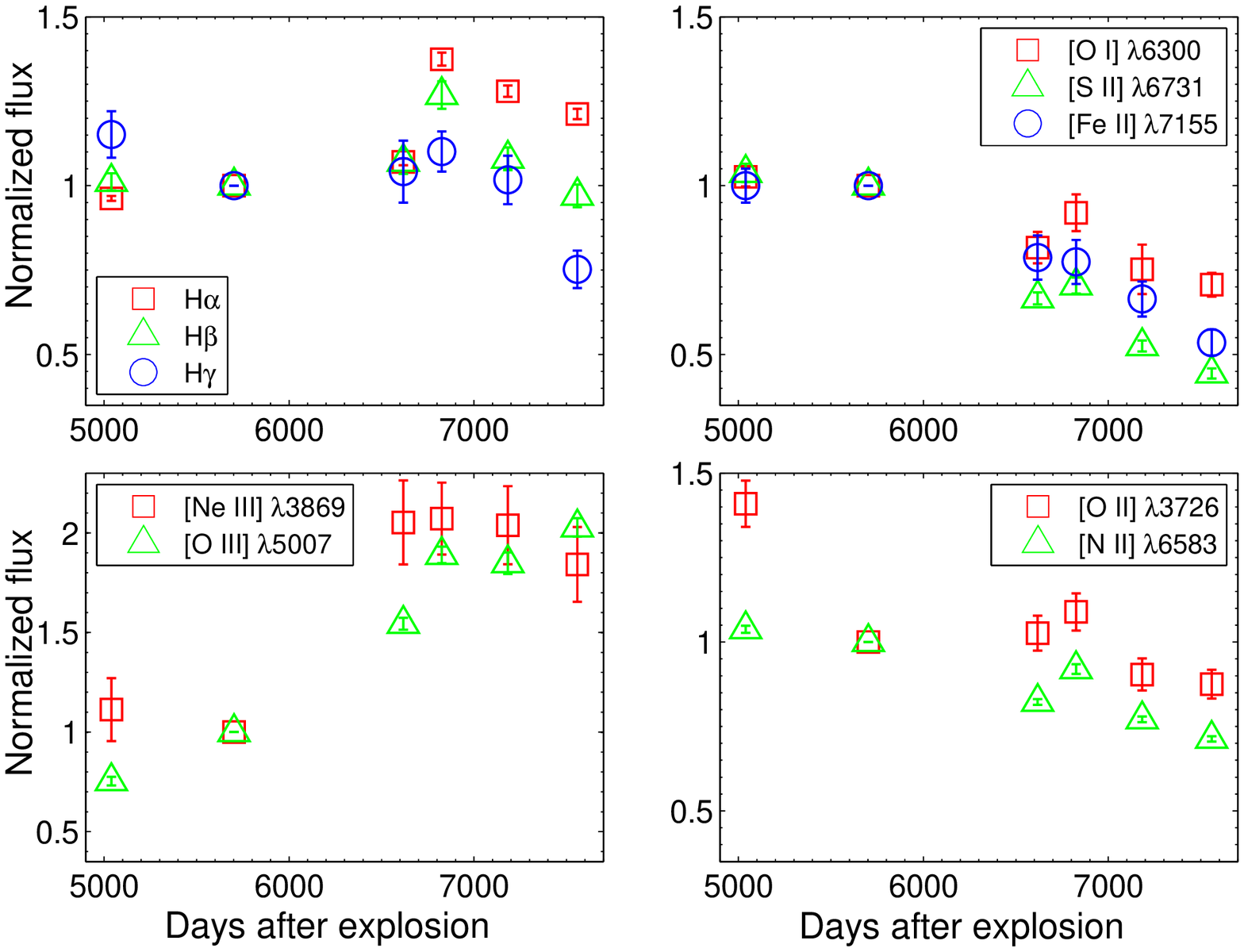}}
\caption{Upper panels: Fluxes for the unshocked, narrow component from the
northern part of the ring, normalized to the flux at Oct. 2002 (day 5703).
Lower panels: Fluxes for the southern part of the ring.}
\label{fig:normflux_vs_time_narrow}
\end{figure}

Figure \ref{fig:normflux_vs_time_narrow} shows the flux evolution of
the narrow lines for the northern and southern parts of the ER (see Tables \ref{tab:narrowlines2}--\ref{tab:narrowlines7}). In order to compare the
trends of the different lines more clearly we have normalized these to the flux
on October 2002 (Epoch 3). Because of the low S/N, we
do not include Epoch 1 in the analysis.

The Balmer line fluxes appear to be constant in time up to day $\sim 6800$, but
after this epoch there is a clear decline. The fluxes from the
low-ionization [N~II], [O~I], [O~II], [S~II] and [Fe~II] lines roughly follow
this trend. In contrast, the highly ionized [Ne~III] and [O~III] fluxes appear
to increase up to day $\sim 6800$. After that they are constant or slowly
decreasing. We return to this interesting result in
Sect. \ref{sec_disc_narrow}. The fluxes of [Ne~V] $\lambda3426$ have a
large statistical uncertainty, and it is therefore difficult to say
anything regarding the trend.

Except for  [Fe~II], 
it is difficult to draw any conclusions from the other iron lines of higher
ionization stage due to their large statistical uncertainties.

\begin{figure}
\resizebox{\hsize}{!}{\includegraphics{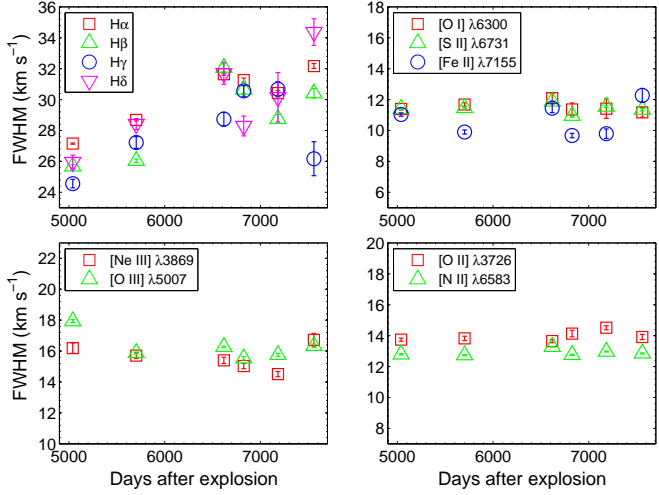}}
\caption{FWHM of the unshocked components as a function of time for the northern part of the ring.}
\label{fig:fwhm_vs_time}
\end{figure}

Figure \ref{fig:fwhm_vs_time} shows the width (FWHM) of the emission lines as a
function of time for the northern part of the ER (see Tables \ref{tab:narrowlines2}--\ref{tab:narrowlines7}). The line widths of the
narrow emission lines depend on the instrumental broadening
($\sim 6\kms$), the bulk motion of the ER and the temperature of the
unshocked gas. The instrumental broadening can be considered
constant from epoch to epoch. The expansion velocity of the ring
should also be fairly constant. However, due to the evolution
of the spatial ring profile, different regions of the ring may be
sampled in the extraction process causing slight changes in the
macroscopic velocity distribution from epoch to epoch. This effect
would in such case broaden the line and become more important for
later epochs. Nevertheless, for the light elements, such as hydrogen and
helium, any differences should be 
small relative to the thermal broadening, and the FWHM from epoch to
epoch should mainly reflect changes in temperatures for these
lines.

As shown in Fig. \ref{fig:fwhm_vs_time}, the widths are fairly constant for
most of the lines. Only the Balmer lines show a tendency of increasing line
widths, which may indicate an increase in temperature of the emitting gas. 
Note, however, that lines with higher ionization stages are slightly wider,
likely due to higher temperature in the emitting gas (see G08).

\begin{figure}
\resizebox{\hsize}{!}{\includegraphics{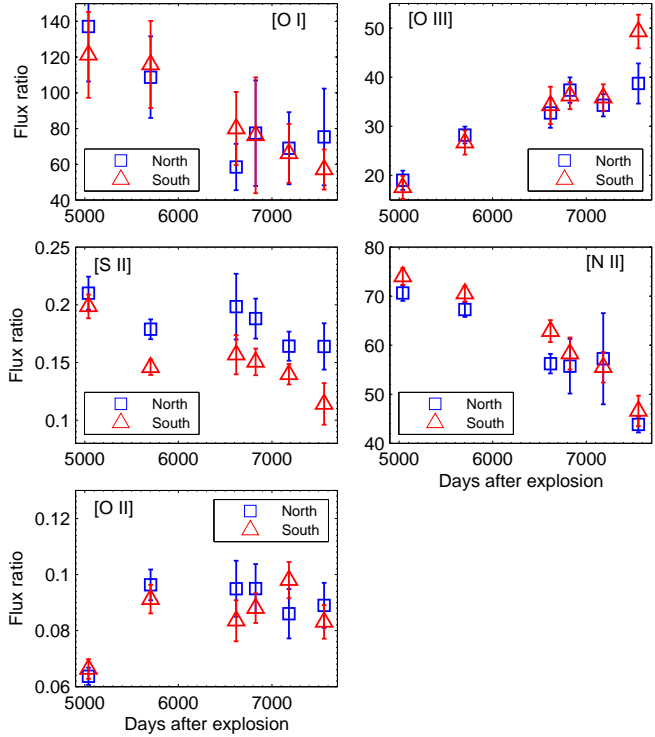}}
\caption{Diagnostic line ratios from the unshocked northern and
southern components.  The ratios shown are [O~I] $(\lambda6300+\lambda6364)/\lambda5577$, [O~III] $(\lambda4959+\lambda5007)/\lambda4363$, [S~II]
$(\lambda4069+\lambda4076)/(\lambda6716+\lambda6731)$, [O~II] $(\lambda\lambda7319-7331)/(\lambda3726+\lambda3729)$ and [N~II] $(\lambda6548+\lambda
6583)/\lambda5755$.}
\label{fig:fluxratios_vs_time_narrow}
\end{figure}

Figure \ref{fig:fluxratios_vs_time_narrow} shows the line ratios for a number
of diagnostic lines for the narrow component.
As can be seen, the flux ratio of
[N~II] $j_{\lambda\lambda6548,6583}/j_{\lambda5755}$ decreases with time, which
implies an increase in temperature.
For [S~II] and [O~I] we see roughly the same behavior as for [N~II].
However, there is no
indication of an increase of the line widths. Instead, the widths appear
rather constant (Fig. \ref{fig:fwhm_vs_time}).
The [O~II] $j_{\lambda\lambda7319-7331}/
j_{\lambda\lambda3726,3729}$ ratio, on the other hand, is rather constant
except for a lower value at Epoch 2 than at later epochs.
By contrast, the [O~III] ratio $j_{\lambda\lambda4959,5007}/j_{\lambda4363}$
increases with time. In general, there are only modest differences between the
flux ratios at each epoch for the northern and southern sides. We return to a
discussion of the narrow lines in Section \ref{sec_disc_narrow}.

\subsection{Evolution of the shocked components}

\subsubsection{Line fluxes}

\begin{figure}
\resizebox{\hsize}{!}
{\includegraphics[scale=0.1]{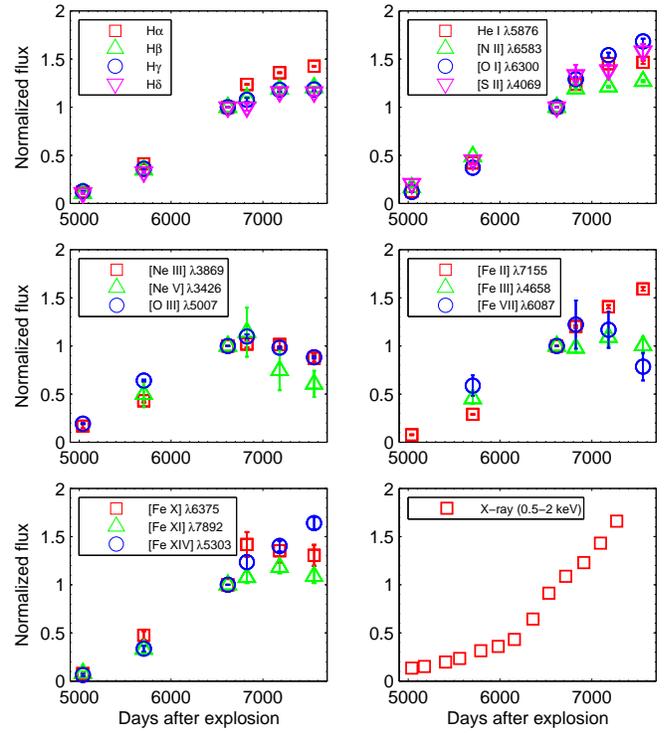}}
\caption{Fluxes for the shocked component from the northern part of
the ring for selected lines, normalized to the flux at Apr. 2005 (day
6618). The lower right panel shows the soft X-ray flux normalized to
the same date \citep[from][]{park07}.}
\label{fig:normflux_vs_time_broad_n}
\end{figure}

\begin{figure}
\resizebox{\hsize}{!}
{\includegraphics[scale=0.445]{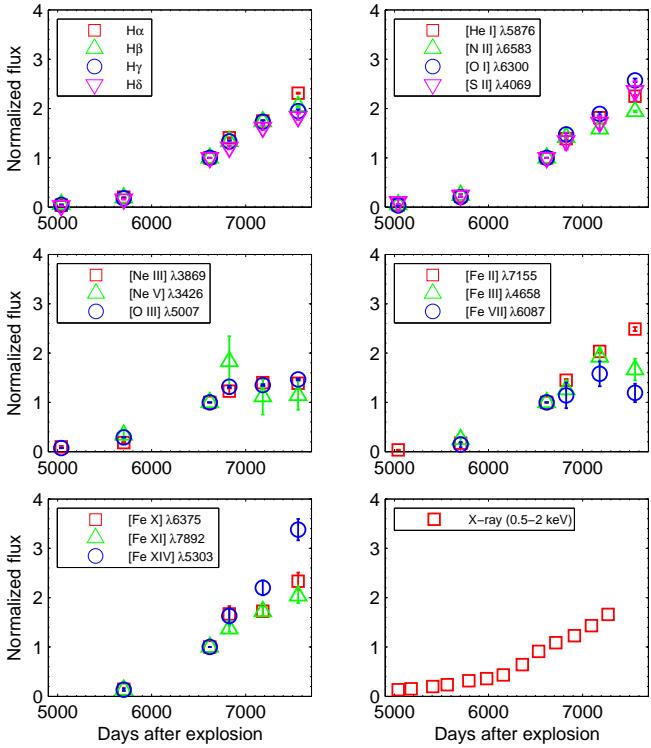}}
\caption{Same as Fig. \ref{fig:normflux_vs_time_broad_n}, but for the
southern part of the ring. Note the change of flux scale compared to
Fig. \ref{fig:normflux_vs_time_broad_n}.}
\label{fig:normflux_vs_time_broad_s}
\end{figure}

The shocked gas is dominated by emission from the hot spots.
A complete list of lines from the shocked component is given in Table 2 in G08.
The fluxes of a selection of representative lines from the
intermediate velocity component are shown in Fig.
\ref{fig:normflux_vs_time_broad_n} and Fig.
\ref{fig:normflux_vs_time_broad_s} for the northern and southern components,
respectively.  As with the narrow lines, we have
normalized the fluxes, but in this case to Epoch 4 (2005 April) to
facilitate a comparison of the different lines more directly. The
absolute values of the fluxes  can be
found in Tables \ref{tab:broadlines_n} and \ref{tab:broadlines_s} in the
Appendix. For comparison we
also show the evolution of the integrated soft X-ray flux in the 0.5 -- 2 keV
range from \citet{park07}. 

As was already found in \citet{Gro06} for the high ionization lines,
the line emission from the shocked gas increases rapidly as more ring
gas is swept up by the shock. Here we confirm that this is a general
feature for all lines. The rate of increase is, however, different for
the different lines (Fig.  \ref{fig:normflux_vs_time_broad_n}). The
low ionization lines of He I, [N~II], [O~I], [S~II] and [Fe~II]
increase similar to each other and to the Balmer lines. This is also
the case for the [Fe~XIV] line. The intermediate ionization lines,
[O~III], [Ne~III], [Ne~V], [Fe~III] and [Fe~VII], however, all show a break
between day $6500$ and day $7000$.
This is probably also the case for the [Fe~X]
and [Fe~XI] lines, although this result is close to the systematic
uncertainties. The same pattern is seen
for the southern component (Fig. \ref{fig:normflux_vs_time_broad_s}).
The Balmer decrement, as seen in H$\alpha$, H$\beta$, H$\gamma$, and H$\delta$,
is constant within the systematic errors.  

\begin{figure}
\resizebox{\hsize}{!}{\includegraphics{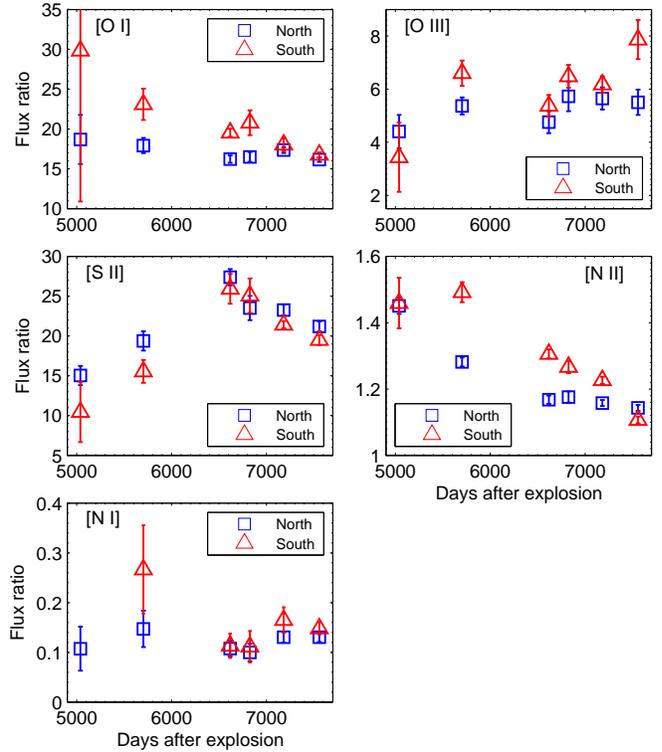}}
\caption{Diagnostic line ratios from the shocked northern and
southern components.  The ratios shown are [O~I] $(\lambda6300+\lambda6364)/\lambda5577$, [O~III] $(\lambda4959+\lambda5007)/\lambda4363$, [S~II]
$(\lambda4069+\lambda4076)/(\lambda6716+\lambda6731)$, [N~I] $(\lambda
5198+\lambda5200)/\lambda3467$ and [N~II] $(\lambda6548+\lambda
6583)/\lambda5755$. Note that the [O~III] $\lambda4363$ line profiles have been
deblended from [Fe~II] $\lambda4358$ which makes the [O~III] flux ratios
somewhat more uncertain.}
\label{fig:fluxratios_vs_time_broad}
\end{figure}

The flux ratios for the shocked components, plotted in
Fig. \ref{fig:fluxratios_vs_time_broad}, are useful diagnostics for
the shocked gas, being sensitive to both electron
density and temperature. In G08 we found that the densities are consistent with
the compressions from radiative shocks. The electron density in the [O~III]
region was constrained to be in the range $\sim10^6-10^7\cc$ and with a
temperature in the interval $\sim(1-4)\EE4\KK$ for both the northern and
southern parts of the ring \citep[see also][]{pun02}. 

For [S~II] we find a density of  $n_e\ga 10^7 \cc$ and a
temperature of the order of $T_e\sim5000\KK$ in this region, while [O~I] has a
somewhat higher temperature, $T_e\sim8000\KK$, and [N~II] implies a marginally
higher temperature than for [O~I].
This result is confirmed by continued observations, but we also see some
evolution in the line ratios in Fig. \ref{fig:fluxratios_vs_time_broad},
consistent with an increasing temperature for [S~II],
[N~II] and [O~I].

The [O~III] flux ratio
$j_{\lambda\lambda4959,5007}/j_{\lambda4363}$ is temperature
sensitive for the considered density interval. The data
suggest that the emission comes from a region with an electron
density of a few times $10^6\ccm$ and a temperature around
$15,000-20,000\KK$. It is difficult to find a clear trend for the
evolution of the flux ratio, perhaps because the line
[O~III]~$\wl4363$ is blended with [Fe~II]$~\lambda4358$, and as a
consequence has a relatively large uncertainty in flux.

\subsubsection{Line profiles}

\begin{figure*}
\resizebox{\hsize}{!}{\includegraphics{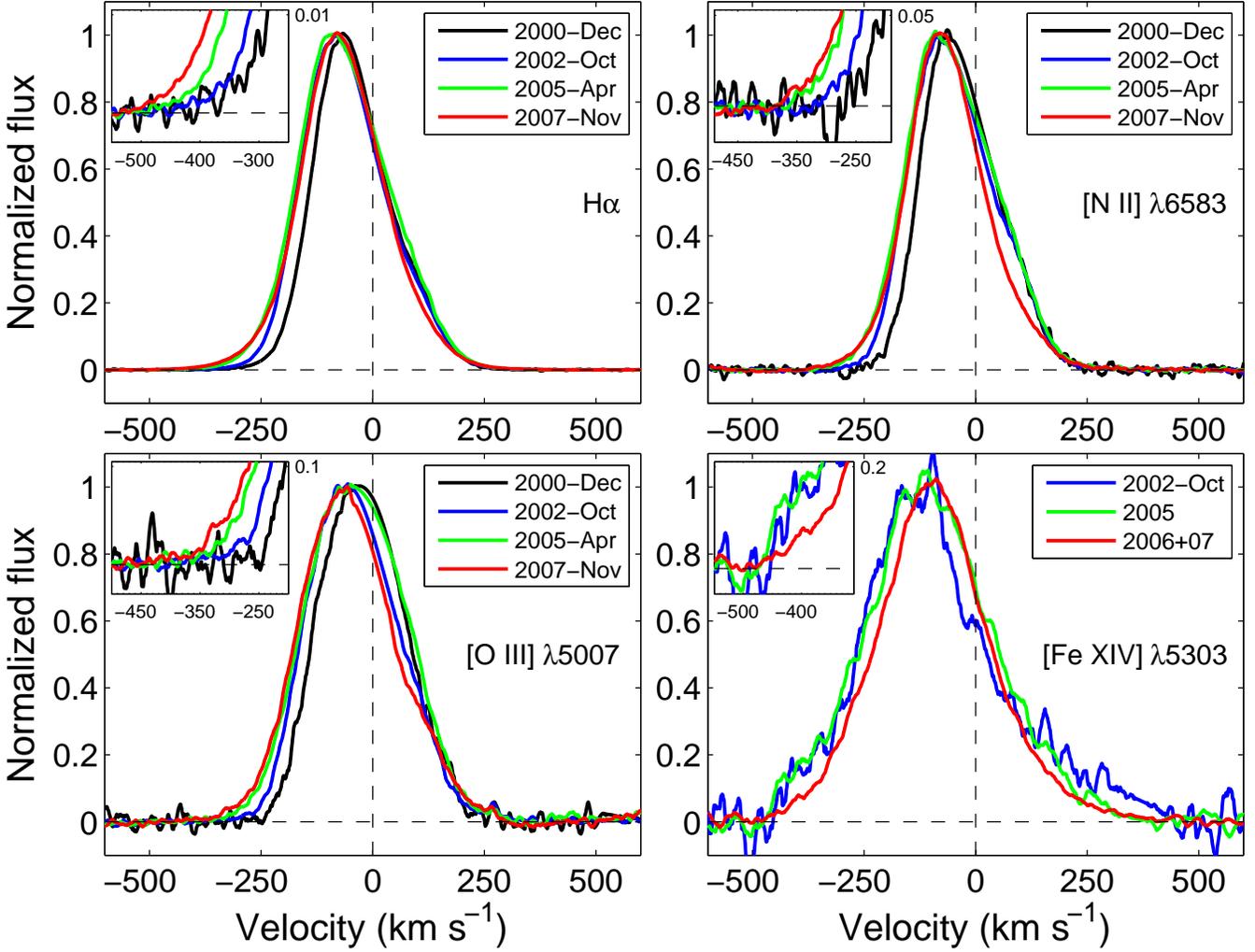}}
\caption{Evolution of the H$\alpha$, [N II] $\wl 6583$, [O III] $\wl
  5007$, and [Fe XIV] $\wl 5303$ profiles from the shocked gas at the
  northern part of the ring. The insert in each panel show the blue
  wings of the lines to more clearly show the extent of the high
  velocity tail in each line. To increase the S/N for the [Fe
    XIV] $\wl 5303$ line we have added observations from 2006 --
  2007. The zero velocity corresponds to the rest frame of the northern part
  of the ring.}
\label{fig:evol_profiles}
\end{figure*}

Figure \ref{fig:evol_profiles} illustrates the evolution of the
H$\alpha$, [N II] $\wl 6583$, [O III] $\wl 5007$, and [Fe XIV] $\wl
5303$ line profiles from Epochs 2 to 7.  These have been
selected to represent a broad range of ionization stages, as well as a
maximum S/N. We have subtracted the underlying broad emission
of H$\alpha$ from the reverse shock to determine the profiles of the
emission from the shocked gas for H$\alpha$ and [N II]~$\wl 6583$,
and for the other lines
the continuum. Any blends have been subtracted using the procedure in
G08. To facilitate a comparison between the different dates we have
for this figure normalized the flux to the peak flux of the line
profile.

\begin{figure}
\resizebox{\hsize}{!}
{\includegraphics[scale=0.1]{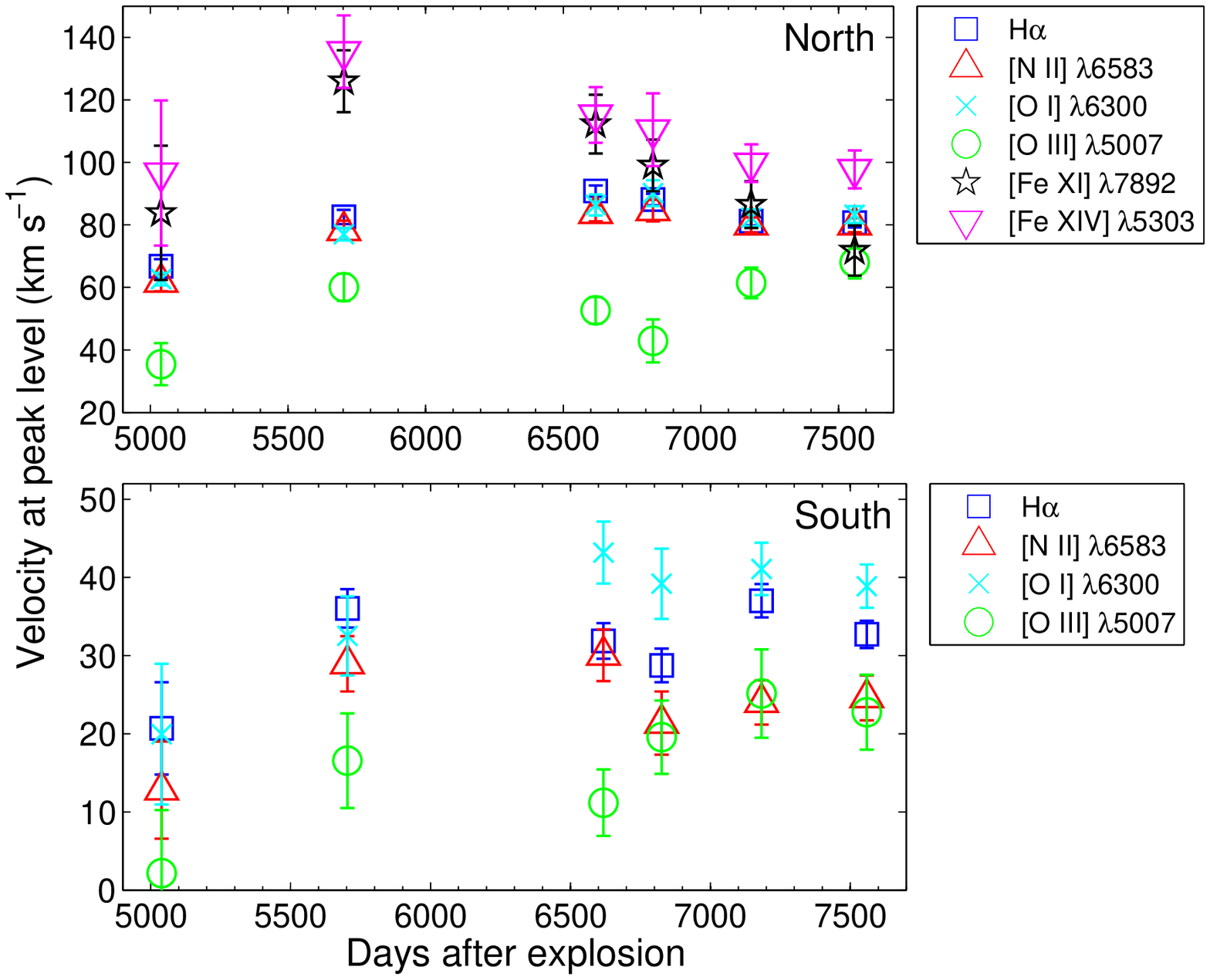}}
{\includegraphics[scale=0.5]{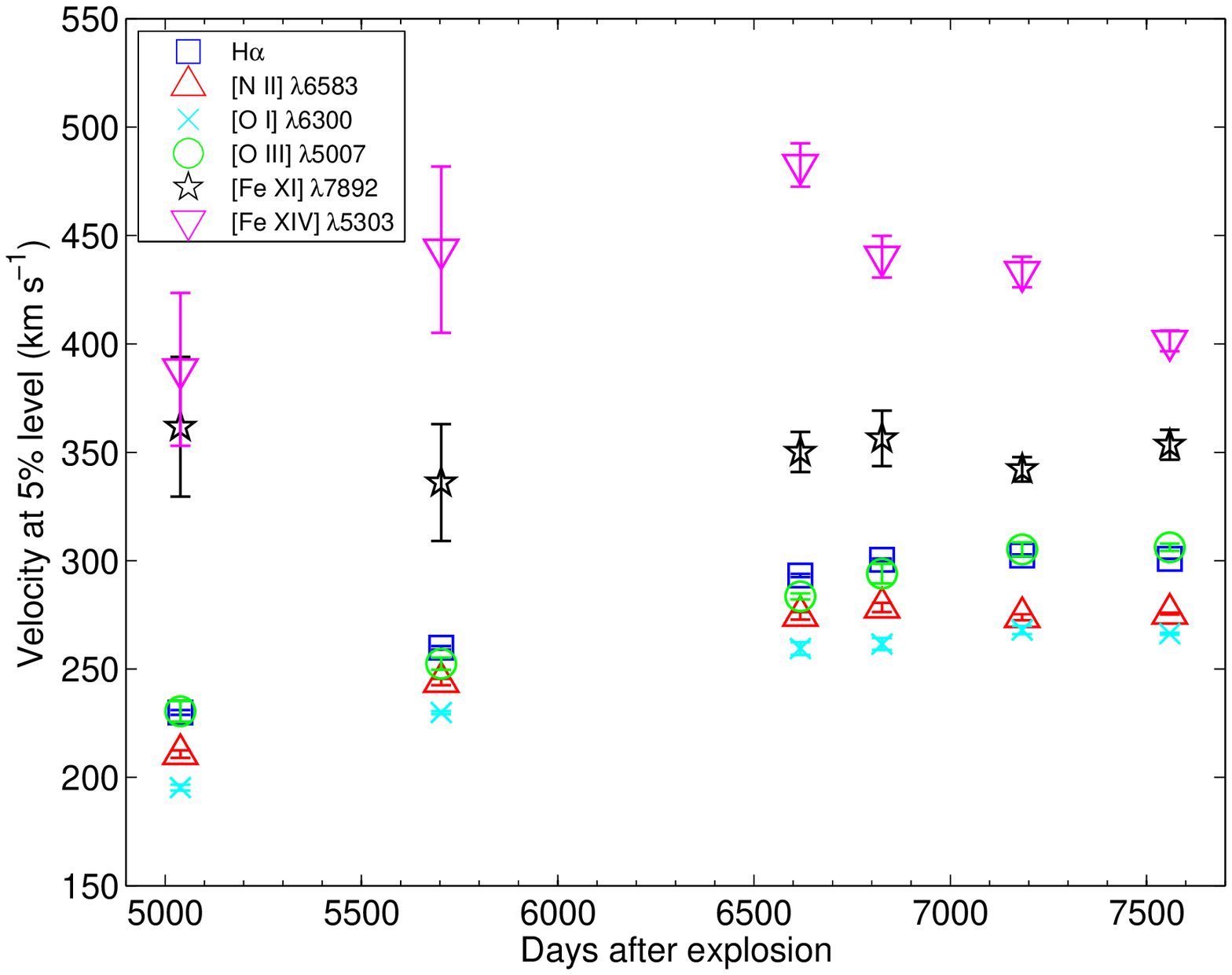}}
\caption{Upper panels: Velocities of the peak of the lines for the shocked
component. Lower panel: Velocities at 5\% flux level for the shocked component
from the northern part of the ring.}
\label{fig:linevel_vs_time}
\end{figure}

Although the whole line profile evolves from epoch to epoch, the
figures reveal that between Epochs 2 to 7 the largest increase
in flux takes place for the blueshifted emission. In particular, the
high velocity end of the blue wing shows a steady increase in
velocity with time.
In Fig. \ref{fig:linevel_vs_time} we show the peak velocity and the velocity
at a level of 5 \% of the peak
flux as a function of time for the different lines. The 5 \% level has been
chosen in order
to facilitate a comparison between the low ionization lines, like H$\alpha$,
with the high ionization coronal lines, like [Fe~XIV] (in the same way as in G08). The latter has a considerably noisier profile due to the low flux
(only $\lsim 1$ \% of H$\alpha$) and disappears in the noise at levels
less than $\sim 5 \%$ of the peak emission. We discuss the velocity at lower
levels in Sect. \ref{sec_disc_interm}.

The peak velocity in Fig. \ref{fig:linevel_vs_time} only marginally evolves
with time in both the southern and northern regions. There
are, however, noticeable differences between the velocity values for the different
lines. The low ionization lines, H$\alpha$, [O~I] and [N~II], have all
similar velocities, while the [O~III] and [Fe~XIV] lines have lower and
higher peak velocities, respectively, showing that contribution
to the different lines probably come from different regions covered by the
slit.

In addition, there are clear differences in peak velocities of the lines at the
northern part of the ring compared to those in the southern part
(Fig. \ref{fig:linevel_vs_time}). The velocities are considerably higher on the
northern part which is likely due to a line-of-sight projection effect
(see Sect. \ref{sec_disc_interm}).
In G08 we compared these peak velocities with the velocities
obtained by the VLT/SINFONI instrument, which has an adaptive optics
supported integral field spectrograph \citep{kjaer07}. While the
spatial resolution for SINFONI is much higher than for our UVES observations,
the spectral resolution is low, $67 - 150 \kms$ in the different
bands, and only gives an average, bulk velocity at each point on the
ring. As found in G08, there is good agreement at the epoch of the
first SINFONI observation in November 2004.

Turning now to the maximum velocity in
Fig. \ref{fig:linevel_vs_time} we see an interesting
difference between the lines, as well as an evolution in time. While
the low and intermediate ionization lines, H$\alpha$, [O~I],
[N~II] and [O~III], show similar velocities, $250 - 300 \kms$, the
coronal lines have considerably higher velocities, $\sim 350 \kms$ for
[Fe~X] and $\sim 450 \kms$ for [Fe~XIV]. Again, this shows that the
two sets of lines come from regions of quite different physical
conditions.  We will discuss this in more detail in
Sect. \ref{sec_disc_interm}.
There is an indication that the coronal line profiles, such as [Fe~XI-XIV],
become less blueshifted
over time for the northern part of the ring. The maximum velocity (FWZI)
does, however, not decrease with time, as seen in Fig. \ref{fig:evol_profiles}.

\section{Discussion}
\label{sec:discussion}

In this section we discuss the implications of the results in Sect.
\ref{sec:analysis}. The general picture we here have in mind for the optical
emission is that of a radiative shock, moving into the dense ER, with a
pre-shock density of $3\EE3-3\EE4\cc$ and a temperature of
$\sim(1-4)\EE4\KK$ (see below).
The narrow line emission comes from this gas.
The soft X-rays, as well as the coronal
lines, then come from the immediate post-shock gas, with a temperature of
$\sim(1-3)\EE6\KK$ and a density four times that of the pre-shock density.
As the gas cools to $\sim10^4\KK$, the density increases by a similar factor to
$10^6-10^7\cc$.
This is where most of the optical, intermediate velocity lines arise. We now
discuss these regions, one by one.

\subsection{The narrow, unshocked lines}
\label{sec_disc_narrow}
In connection with the shock breakout the rings were ionized by the
soft X-rays during the first hour \citep{LF96}. After this they have
recombined and cooled at a rate determined mainly by the density of
the different parts of the rings. However, because of the strong X-ray
flux from the shocks propagating into the dense blobs one expects 
re-heating and re-ionization of this gas, as well as gas in the inner ring and
surrounding gas that was never ionized by the supernova flash. The decrease in
the line fluxes after the flash ionization is therefore expected to turn into
an increase, or at least a slower decay, when preionization by the shocks
starts to dominate the conditions in the ring. The evolution of the narrow
components of the different lines is therefore of great interest. 

As shown in Fig. \ref{fig:normflux_vs_time_narrow}, we do indeed find that the
evolution of the fluxes of the narrow lines is
different for different lines. The [O~I], [O~II], [N~II], [S~II] and
[Fe~II] lines all decrease with time
(Fig. \ref{fig:normflux_vs_time_narrow}). These lines are all
collisionally excited and therefore sensitive to both temperature and
degree of ionization. The hydrogen lines arise mainly by recombination
and therefore decay slower than the collisionally excited lines, as is
observed.

The most interesting result is, however, the observed
{\it increase} of the [O~III] and [Ne~III] lines up to day $\sim 6800$. 
After this epoch the fluxes of these lines become constant or
decrease. This behavior
is consistent with calculations of the shock structure
of a radiative shock.  Soft X-rays from the shock ionize
and heat the pre-shock gas to a temperature of $(2-3) \times 10^4$ K
immediately in front of the shock \citep[][]{allen08}. The absorbed X-rays are
there re-emitted as UV and optical lines. The interesting point is now that
these shock calculations show that the [O III] $\wll 4959, 5007$ lines
are by far the strongest optical lines from this region. The fact that
these lines show an increase in the fluxes is
therefore consistent with the shock models, and
strongly indicates that the radiation from the shock interaction started to
affect the unshocked ring already several years ago.

The levelling off of the [O~III] and [Ne~III] lines
between day $6500$ and day $7000$
may be a result of several factors. One possibility is that all
the gas in the ring with high enough density to be visible (i.e.,
$n_e\gsim10^3 \ccm$) has been ionized by the X-rays, which will result in a
nearly constant flux after this epoch. It could, of course, also be
the result of a decreasing X-ray flux from the shocks, or finally from
sweeping-up of the dense gas by the shocks.

Similarly, the fast decrease of the low ionization lines may either be a
result of ionization or sweeping-up of the unshocked gas. Future HST imaging
may distinguish between these scenarios.

A more detailed picture of the physical conditions of the narrow-line
emitting gas can be obtained from line ratios (see
Fig. \ref{fig:fluxratios_vs_time_narrow}).  The declining diagnostic
ratios of [N~II], [O~I] and [S~II] all suggest an increasing
temperature in this zone, which may be indicative of pre-ionization.  
 
For the [O~II] and [O~III] lines we see a rather dramatic change between
Epochs 2 and 3 compared to at later epochs. The low ratios for
[O~III] $j_{\lambda\lambda4959,5007}/j_{\lambda4363}$ at Epoch 2 indicate much
higher temperatures in both the northern and southern parts of the ring than at
later epochs. Using nebular analysis as in G08, the [O~III] temperature appears
to decline from $\sim 5\EE4$~K to $\sim 2.5\EE4$~K between Epoch 2 and 3, and
further down to $\sim 2\EE4$~K at later epochs. The Epoch 2 temperature is
clearly much higher than expected from a shock precursor \citep[cf. above, see also][]{allen08}.

The most likely reason for this is that the narrow-line emission at the
earliest epochs comes mainly from low-density gas that was flash-ionized by the
supernova.
\citet[][ and paper in preparation]{mattila03} show that the density of the
flash-ionized gas dominating the [O~III] emission at epochs around 5000 days is
$\sim 1\EE3 \cc$. The [O~III] temperature in these models is $\sim 5\EE4$~K,
i.e., close to what we observe. From an inspection of the HST images
(see Fig. \ref{fig:ringimage_hst}) it is also clear that by day
$\sim 5000$ rather few spots on the ring have 
been shocked, and that most of the emission within the UVES slit at Epoch 2
comes from the diffuse, rather evenly distributed, emission between the blobs.
This is even more pronounced in [O~III] than the H$\alpha$+[N~II] emission
shown in Fig. \ref{fig:ringimage_hst}. Note also that only a fraction of the
emission from the blobs seen in the HST images stems from narrow-line emitting
shock precursors, as the HST images cannot distinguish between that emission
and the line emission from the shocked gas. 

At Epoch 2, we estimate that
$\lsim 30$\% of the narrow-line [O~III]~$\lambda\lambda4959,5007$ emission
comes from shock precursors, and that the rest comes from the diffuse
low-density gas. For [O~III]~$\lambda4363$ the fraction would then be
$\lsim 15$\% to obtain the observed
$j_{\lambda\lambda4959,5007}/j_{\lambda4363}$ ratio. However, already by
Epoch 3, the narrow [O~III]~$\lambda\lambda4959,5007$ emission within the UVES
slit should be dominated by shock precursors, whereas the narrow
[O~III]~$\lambda4363$ would still be ``contaminated" and perhaps even dominated
by the $\sim 1\EE3 \cc$ component which would then still have a temperature
of $\sim (4-5)\EE4$~K, according to the models discussed in
\citet[][]{mattila03}. The higher temperature for [O~III]~$\lambda4363$ is
directly supported by the larger line width for that line seen at Epoch 3,
compared to the other two [O~III] lines (G08).

This two component model is also consistent with the evolution of the fluxes
from the narrow and intermediate velocity components.
The increase in the flux of the intermediate-velocity component of
[O~III]~$\lambda\lambda4959,5007$ should correlate with the narrow-line flux
from the precursor. Between Epochs 2 and 3, the former increases by a factor
$\sim 3.4$ for the northern part of the ring, while the narrow velocity
component only increases by a factor of $\sim 1.3$.
However, after Epoch 3, when the precursor emission is likely to dominate
[O~III], the increase of the narrow component is similar to the intermediate
velocity flux.

The extent of the pre-ionized zone depends on the number of ionizing photons
from the cooling shock, which should scale as the total flux
${1\over2}\rho V_s^3$. The number of recombinations per unit area is
$\alpha_B n_e^2 \Delta x({\rm HII})$, where $\Delta x({\rm HII})$ is the
thickness of the ionized region and $\alpha_B$ is the Case B recombination
rate. Therefore, $\Delta x({\rm HII})\propto V_s^3/\alpha_B n_e^2$. This
describes well the shock calculations by \citet[][]{allen08}, which gives the
normalization, and we find

\begin{equation}
\Delta x({\rm HII})\approx 1.4\EE{16}\left({V_s\over 200\kms}\right)^3\left({n_e\over 10^4\cc}\right)^{-1}{\rm~cm.}
\end{equation}

The pre-ionized region is therefore sensitive to both the shock velocity, which
probably spans a range $100-400\kms$, as well as the density. It is clear that
a large fraction of the ring may be pre-ionized by the shocks, as well as
material outside of the ER, which is likely to have a lower density.

\subsection{Shocked intermediate component}
\label{sec_disc_interm}
From Fig. \ref{fig:normflux_vs_time_broad_n} and Fig. \ref{fig:normflux_vs_time_broad_s} we found that the lines from the shocked gas increase by a factor of
$\sim 4$ from day 5703 to around day 7000 for the
northern ring, and by an even larger factor for the southern
ring. This is similar to that of the soft X-rays, and for comparison we
show the evolution of the soft X-rays from \citet{park07} as the lower
right panels in Fig. \ref{fig:normflux_vs_time_broad_n} and Fig. \ref{fig:normflux_vs_time_broad_s}. Note that the X-ray
flux is that
of the total ring flux, while our fluxes represent two different
regions. We find nevertheless that the optical flux and soft X-ray flux
have a very similar evolution, and that they therefore are likely to
come from the same region.

One of the most important results in our study of the shocked,
intermediate velocity component is the increase in the widths (FWZI)
of the low ionization lines (Fig. \ref{fig:evol_profiles}). With the
exception of the coronal lines, the optical line emission must
originate in radiative shocks. The widths of the shocked emission lines
are therefore expected to increase with time as faster shocks become
radiative.  However, the line profiles only reflect the projected
shock velocities along the line of sight, and consequently, the FWZI
velocity is only a lower limit to the maximum velocities of the
radiative shocks driven into the protrusions. Nevertheless, if
conditions are similar, a steady increase in the line widths with time
is therefore a strong indication that gas from shocks with
increasingly higher shock velocity did have time to cool. This is
further strengthened by the fact that in G08 we found that the line
widths of the low ionization lines in 2002 were considerably smaller
than those of the coronal lines. Now, in our last observations in 2007
(Epoch 7) the maximum blueward velocities of the low ionization lines
are approaching those of the coronal lines (Fig. \ref{fig:linevel_vs_time}).

\begin{figure}
\resizebox{\hsize}{!}{\includegraphics{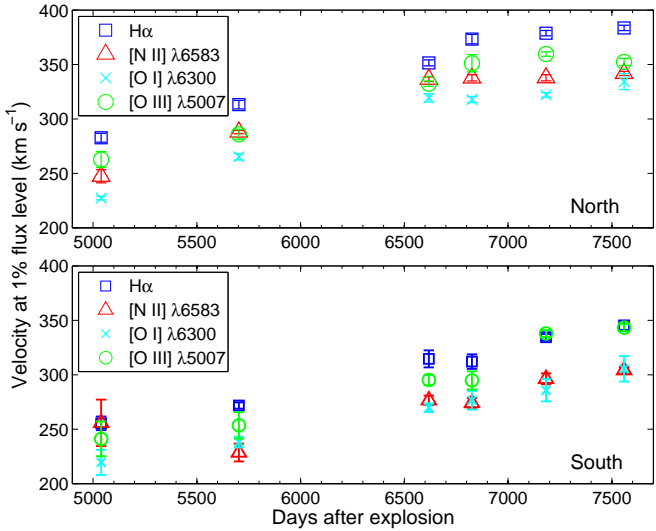}}
\caption{Velocities at 1\% flux level for the shocked gas from the northern and
southern components.}
\label{fig:maxvel_vs_time_1p}
\end{figure}

The maximum velocities estimated in
Fig. \ref{fig:linevel_vs_time} are based on the velocity at 5 \%
of the peak flux. This was mainly motivated by the requirement to compare
the coronal lines with the low ionization lines. Because of the
high S/N of the strong lines we can detect the line wings to
considerably lower levels to get a better estimate of the true
velocity extent of the lines. In Fig. \ref{fig:maxvel_vs_time_1p} we
show the maximum velocity from the northern and southern parts of the
ring for H$\alpha$, [N~II], [O~I] and [O~III], now at a 1 \% level of the peak
flux. We here see the same increase in maximum velocity with
time as in Fig. \ref{fig:linevel_vs_time}, although the velocities are
considerably higher.
From Fig. \ref{fig:evol_profiles} we see that H$\alpha$ can in fact
be traced to $\sim 500 \kms$ in the last spectrum.
The same is true for the southern part of the ring.

As can be seen from Figs. \ref{fig:linevel_vs_time} and
\ref{fig:maxvel_vs_time_1p}, the velocities from the southern part
of the ring are consistently lower by $\sim 40 \kms$, but do also show a
similar
increase with time as those from the northern part. The lower
velocities may either be a projection effect, an effect of the later
impact in this region, or a higher density.
Even though there is clearly a time delay between the northern and southern
parts of the ring ($\sim300$~days), the most plausible explanation is a
difference in the projection of velocities between the two regions. The flux
from the northern part mainly comes from ``hot spots'' between
$\sim0^\circ-50^\circ$
while spots between $\sim230^\circ-250^\circ$ dominate the flux from the
southern part (see Fig. \ref{fig:ringimage_hst}).
This asymmetric distribution of ``hot spots'' between the northern and southern
parts of the ring could well account for a difference in projection such that
the line-of-sight velocities from the southern part are consistently lower
than the velocities from the northern part, consistent with the observed line
profiles. This is also supported by the similar difference in peak velocity,
which is $40-50 \kms$ \citep[see also][]{kjaer07}.

For a shock to become radiative it must cool from its postshock
temperature to $\sim 10^4$ K from the time it was shocked. From shock
calculations we found that the time it
takes for a shock to become radiative, i.e., the cooling time, $t_{\rm
cool}$, is related to the preshock density and the shock velocity as
\begin{equation}
t_{\mathrm{cool}} \approx 8.3 \left({n_{\mathrm{spot}} \over
10^4\cc}\right)^{-1} \left({V_s\over 300\kms}\right)^{3.4} \ \rm years
\label{eq:cooltime}
\end{equation}
\noindent 
\citep{Gro06}, where $n_{\mathrm{spot}}$ is the pre-shock density of the ring.

At Epoch 3 (October 2002) the shocks had $\sim 7 \rm{~years}$ to cool since
first
impact \citep[taken as $1995$; see ][]{law00}, and from the FWZI of the
line profiles, we estimate that shocks with velocities $\lsim 300\kms$
have had enough time to cool. This places a lower limit on the
preshock density from Eq. (\ref{eq:cooltime}) and we estimate
$n_{\mathrm{spot}}\gsim1.2\times10^4\cc$.  At our last epoch the maximum
velocity of H$\alpha$ is $\sim 500 \kms$. With this
velocity Eq. (\ref{eq:cooltime}) implies a density of
$n_{\mathrm{spot}}\gsim 2.7\times10^4\cc$.

These densities can be compared to the densities derived from the
analysis of the ring during the first few years after
outbreak. \citet{LF96} find that the light curves of the ring emission
require a range in densities from $6\EE3 \cc$ to $3.3\EE4 \cc$.
Further, \citet[][ and paper in preparation]{mattila03} show that
later observations of the narrow optical
lines require densities down to $\sim10^3\cc$.
This density agrees with the limit on the ionization time
$n_e t_i\gsim 10^{12}\cc {\rm~s}$ found from the soft X--rays by
\citet{park07}. Using this, and the assumed impact times between $4000-6000$
days after the explosion, \citet{dwek08} find a range of $(0.3-1.4)\EE3\cc$,
which should be seen as a lower limit to the density of the pre-shock gas.
It is likely that we now see the densest blobs cooling down first, and the
agreement with the densities determined for the ring before the
collision makes this scenario likely. As time goes, gas shocked by
higher velocity shocks and/or lower densities will gradually cool
down.

In G08 we showed that the line width of the low ionization lines were
considerably smaller than the coronal lines in 2002 (see also
Fig. \ref{fig:linevel_vs_time}). A likely interpretation is that
part, or all, of the gas emitting the coronal lines at that time had not
yet had time to cool and be seen as optical low ionization emission
lines. The fact that the H$\alpha$ velocities from the northern part are
now comparable to those of the coronal lines is consistent with
the estimates of the cooling time above.
In Fig. \ref{fig:normflux_vs_time_broad_n} we see that all lines increased in
flux up to day $\sim6800$. After
this epoch there are, however, clear differences between the various lines.
While low ionization lines, such as [O~I], [N~II], [S~II] and [Fe~II] continue
to increase, the intermediate ionization lines of [O~III] and [Ne~III] show a
break, and are consistent with being either constant or even decreasing
in flux.

The evolution of the [O~II] and [O~III] line fluxes may be understood as a
result of radiative shocks, which have not
yet had time to cool completely. In such a shock, gas with a given ionization
stage will first increase its emission measure as gas is recombining. After
some time this gas will recombine to the next lower ionization stage and an
equilibrium will be set up, where the emission measure of the ion will be
constant. This will continue to lower and lower ionization stages. At the
lowest ionization stage, this will, however stop and its emission measure will
instead continously increase as long as more cooling gas is accumulating and
the cooling time at this temperature is long compared to the age.
Applied to oxygen, this means that [O~I], [O~II] and [O~III] will all
initially increase. After some time, [O~II] and [O~III] will reach a steady
state, while [O~I] will continue to increase, just as observed.
The same applies
to sulphur and iron, although it will in these cases stop at [Fe~II] and
[S~II], because of photoionization of Fe~I and S~I. The lower ionization lines
will therefore continue to increase as long as new gas is shocked and the
cooling time is long compared to the age of the shock. This picture has been
confirmed with numerical calculations.

[Fe~XIV] and other coronal lines are special in that they may come from
both radiative and adiabatic shocks, as long as the temperature behind the
shock is $\lsim 5\EE6\KK$. At least part of their emission may therefore be
from non-radiative shocks, which is consistent with the more extended line
profiles, especially at early epochs.

In a recent paper \citet{dewey08} find that (adiabatic) shock model fits of the X-ray line
fluxes indicate a bimodial distribution of temperature. The dominant component
has a temperature of $\sim 0.55$ keV, nearly constant between 2004
and 2007. The hard component, on the other hand, decreases from $\sim
2.7$ keV in 2004 to $\sim 1.9$ keV in 2007. Although uncertain, the
FWHM of the lines correspond to a shock velocity of
$200-450\kms$. This is comparable to our optical velocities and
strengthens the identification of the soft X-ray component with the
optical lines, as was already indicated from the flux evolution. The
harder X-ray component does, however, not have any corresponding
optical component. Further, a temperature of $\sim 2.3$ keV corresponds to a
shock velocity of $\sim 1400 \kms$. Such a component should be
adiabatic for a very long time. From the fact that such a high velocity is not
seen in the line widths, Dewey et al. suggest that the high velocity component
may originate from a reflected shock, moving back into the previously
shocked gas.

An interesting additional fact is that the Chandra imaging shows an
expansion of the X-ray emitting plasma corresponding to a velocity of
$1400-1500\kms$ \citep[e.g.,][]{park07}. Also the extent of the
optical emission increases with a similar velocity (P. Challis
priv.comm.). In $\sim10$ years the emitting region has therefore
expanded by $\sim3\EE{16}{\rm~cm}$, which is a considerable fraction
of the thickness of the visible ring. As \citet[][]{dewey08} point
out, this expansion is, however, likely to take place in a low density
component, and not in the dense clumps. We expect this expansion to continue as
long as there is low density gas in the vicinity of the clumps.

%___________________________________________________________________

\section{Summary}
\label{sec:summary}

We have presented high spectral resolution UVES observations at
multiple epochs of the inner circumstellar ring of SN 1987A. Due to
the high spectral resolution, we are able to separate the shocked
from the unshocked ring emission, as well as to study
the evolution of the fluxes and line profiles of the different lines.

For the narrow lines we find strong evidence for pre-ionization of the
unshocked ring by the soft X-rays from cooling gas behind the
shocks. This is indicated both by the increasing [O~III] flux and from
the temperature inferred from the low ionization lines [N~II], [O~I]
and [S~II].  The increasing [O~III] line ratios at the early epochs
(until day $\sim5700$) are consistent with emission coming mainly from very low
density
regions ($\sim1\EE3\ccm$) ionized by the SN flash in connection with
the explosion. Later on, however, the emission is dominated by gas
pre-ionized by the X-rays.

For the intermediate velocity low ionization lines, we find evidence
for increasing line widths for the shocked components. This is
to be expected as faster shocks will have enough time to 
become radiative and cool. At the latest epoch the H$\alpha$ line profile
extends to $\sim500\kms$ for both the northern and southern part of
the ring, which is comparable to the maximum velocity of the coronal lines. The
ring collision in SN 1987A is therefore one of the best examples of high
velocity radiative shocks we have. Together with the X-ray observations
these observations therefore offer a unique opportunity to study this
class of shocks in detail.

The difference in shapes for the line profiles between the northern and
southern components cannot solely be explained by a time delay effect, but is
most likely explained by a difference of projection of the shock velocities
caused by the locations of ``hot spots'' around the ring.

We find a correlation of our optical light curves to that of the soft
X-rays. This suggests that most of the optical lines and soft X-rays
arise from the same region. The difference in line widths of the
coronal lines and the low ionization lines at early epochs indicates that
at least some of the coronal emission comes from adiabatic
shocks. Short of day $\sim7000$ we find  that the optical
light curves of the intermediate ionization lines such as [O~III] and
[Ne~III] is levelling off and especially so on the northern part of
the ring. We argue that this is due to the fact that with time the gas
will recombine to lower ionization stages until the lowest stage is
reached.  If, however, the fluxes from the optical lines have started
to decline, just as the narrow optical lines have done, this may be an
indication that the forward shocks have now
started to interact with gas with lower densities. Future observations
will shed more light on this.

%__________________________________________________________________

\begin{acknowledgements}
We are grateful to the referee, Patrice Bouchet, for his constructive
comments,
which have improved the paper and to the observers as well as the staff at
Paranal for performing the observations at ESO/VLT.
This work has been supported by grants from the Swedish Research
Council, the Swedish National Space Board and STScI for the SAINTS project
(GO-11181).

\end{acknowledgements}

%___________________________________________________________________

\clearpage

%----------------Appendix-----------------------------
\appendix
\section{Comparison of fluxes from HST and UVES.}
\label{sec:flux_comp}

Pipeline-processed HST imaging have been obtained by the SAINTS team
(PI: R.P. Kirshner) at similar epochs as for the
UVES spectra. The instruments were WFPC2 with narrow-band filters F502N and
F656N at day 5013 and ACS with filters F502N and F658N
(at days 5795-7229).

In order to compare the HST fluxes with the fluxes of UVES we first converted
the counts per electrons in the HST images to flux density units
(given in$\ergsma$).
Thus, the images were multiplied by the PHOTFLAM header keyword and, in the
case of WFPC2 images, also divided by the exposure time.
The next step was to rotate the images to $30^\circ$ in accordance with the
slit rotation for the UVES spectra.
To account for the moderate seeing of the ground-based UVES exposures
(typically $\sim0\farcs8$) we artificially added this seeing to the HST images
by convolving each image with a two-dimensional Gaussian kernel.

All the pixel columns in the images within $0\farcs4$ from
the center of the SN (the aperture corresponding to the width and location of the UVES slit) were summed. In this way we obtained spatial flux profiles which
correspond to that of the UVES spectra.
The LMC background emission level was estimated by fitting a second-order
polynomial to the regions outside the outer rings. The polynomial was then
subtracted from the total spatial profile.

The HST fluxes could now be extracted in similar way to the UVES spectra by
fitting a two-Gaussian profile to the bimodal spatial ring profiles. The area
under each Gaussian then corresponds to the northern and southern part of the
ring respectively (see G08 for more details).

To compare the fluxes obtained from the HST images with the UVES fluxes, we
performed a bandpass spectrophotometry of the spectra. Hence, we obtained the
HST filter transmission functions by using the IRAF task calcband in the
hst/synphot package and then integrated the UVES fluxes over the normalized
versions of these.

\begin{figure}
\resizebox{\hsize}{!}{\includegraphics{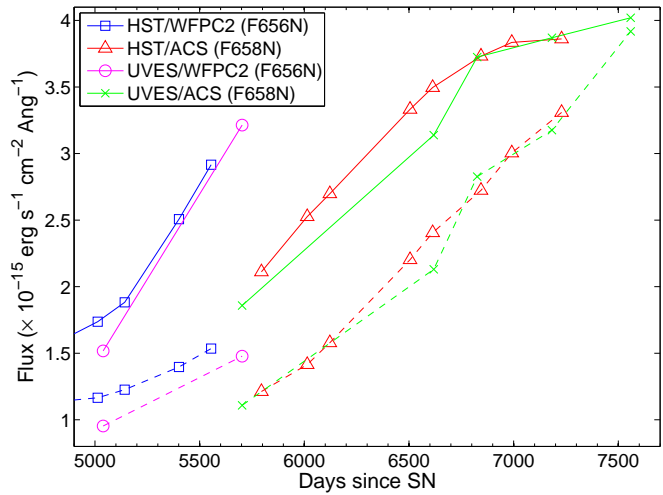}}
\caption{HST narrow-band filter fluxes for WFPC2 and ACS together with the corresponding UVES bandpass fluxes for the F656N and F658N filters for the northern part (solid lines) and the
southern part (dashed lines) of the ring.}
\label{fig:WFPC2f656n_ACSf658n}
\end{figure}

\begin{figure}
\resizebox{\hsize}{!}{\includegraphics{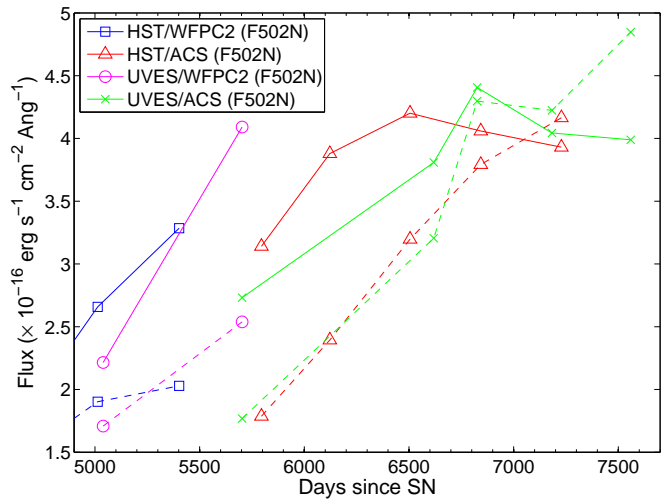}}
\caption{The same as Fig. \ref{fig:WFPC2f656n_ACSf658n} but for WFPC2 and
ACS narrow-band filters F502N.}
\label{fig:WFPC2f502n_ACSf502n}
\end{figure}

The UVES bandpass fluxes have been plotted for different epochs and are
compared with WFPC2 filter F656N and ACS filter F658N images (Fig. \ref{fig:WFPC2f656n_ACSf658n}). In the same way, the UVES
bandpass fluxes are compared with the WFPC2 and ACS filters F502N in Fig.
\ref{fig:WFPC2f502n_ACSf502n}.

In Fig. \ref{fig:WFPC2f656n_ACSf658n} we see that we can directly compare the UVES bandpass fluxes at day 5039 with that obtained by HST/WFPC2 at day 5013. We
find that the UVES flux is $87\%$ of the corresponding HST flux on the northern part of the ring and $82\%$ on southern part. For the ACS fluxes, the largest
deviation occur at Epoch 4 (day 6618). At this epoch the UVES fluxes on the
northern and southern parts of the ring are $90\%$ of that obtained from
HST/ACS at day 6613.

We now turn to Fig. \ref{fig:WFPC2f502n_ACSf502n} for the F502N filter. A comparison between the
UVES fluxes at day 5039 with the HST/WFPC2 fluxes at day 5013 reveals that the
UVES flux is $83\%$ of that of HST on the northern part and $90\%$ on the
southern part of the ring. For the ACS we find that the deviations between the HST
fluxes and the corresponding UVES fluxes are within $\sim15\%$.

From this analysis we estimate that the uncertainty in the UVES absolute
fluxing should be less than $\sim20\%$. In addition, comparing the fluxes from
filter F656N/F658N with those from filter F502N give us a clue about the
accuracy of the relative fluxes between different parts of the spectra and we
find that the uncertainty should be less than $\sim10\%$. The UVES
spectrograph is equipped with three different CCDs (see Sect. \ref{sec:observations}), one in the blue part of
the spectrum ($\sim303-499~\rm{nm}$) and two in the red part
($\sim476-577~\rm{nm}$ and $\sim583-684~\rm{nm}$ respectively for the standard
setting 346+580).
Hence, in this analysis we are only probing the fluxes for two out of the three
CCDs and the uncertainty in the relative fluxes can therefore possibly be
larger for emission lines situated in the blue part of the spectrum. 
The relative fluxes over the whole spectral range have, however, been
checked against UVES spectrophotometric standard star spectra (see Sect. \ref{sec:observations}).

Moreover, to check the robustness of these results we varied
the artificially imposed seeing on the HST images between $0\farcs4$ and
$1\farcs0$. This showed that the total flux covered by the slit is fairly insensitive for this seeing range, and differs by less than a few percent. The redistribution effect of the fluxes from different parts of the ring is, of course, still important as the seeing is varied. The net effect here is that the flux on the northern part of the ring tends to decrease with increasing seeing, while the opposite effect holds for the southern part. This is explained by the fact that the ``hot spots'' are distributed unevenly around the ring (see Fig. \ref{fig:ringimage_hst}) and this effect will therefore vary with time. We find that for
some epochs, the individual fluxes from north and south respectively could
shift by an amount of $\sim10\%$ due to this effect.

In addition, from the HST images we can estimate the relation between the ring
flux encapsulated by the slit and the total flux from the ring. For a simulated
seeing of $0\farcs7$ we found that the factor between total ring flux and the
flux encapsulated by the $0\farcs8$ wide slit was $2.4$ with an accuracy of
$10\%$ for all epochs.

\clearpage

\section{Line fluxes}

\begin{table*}
\caption{Emission lines from the unshocked gas of the ER at Epoch 2.}
\begin{tabular}{llccccccc}
\hline
\hline
&&&North&&&South&&\\
 & Rest wavel.& &$V_{\rm peak}^{\mathrm{c}}$&$V_{\rm FWHM}$&&$V_{\rm peak}$&$V_{\rm FWHM}$&Extinction\\
Emission&$\ang$&Relative flux$^{\mathrm{a}}$&$\kms$&$\kms$&Relative flux$^{\mathrm{a}}$&$\kms$&$\kms$&correction$^{\mathrm{b}}$\\
\hline
\lbrack Ne V\rbrack&3425.86&$19.8\pm2.7$&$282.1\pm1.2$&$20.72\pm0.72$&$17.2\pm2.6$&$294.9\pm1.3$&$18.59\pm1.31$&2.06\\
\lbrack O II\rbrack&3726.03&$162.0\pm4.7$&$283.1\pm0.4$&$13.74\pm0.10$&$152.4\pm5.2$&$293.2\pm0.4$&$14.30\pm0.16$&1.98\\
\lbrack O II\rbrack&3728.82&$97.5\pm5.8$&$280.6\pm0.5$&$12.90\pm0.22$&$101.4\pm6.2$&$290.1\pm0.5$&$15.06\pm0.17$&1.98\\
\lbrack Ne III\rbrack&3868.75&$39.5\pm2.6$&$281.7\pm0.6$&$16.19\pm0.34$&$29.6\pm3.6$&$293.5\pm0.9$&$17.75\pm0.58$&1.94\\
\lbrack S II\rbrack&4068.60&$23.4\pm1.5$&$281.4\pm0.5$&$11.55\pm0.19$&$22.3\pm1.2$&$292.2\pm0.5$&$12.95\pm0.29$&1.90\\
\lbrack S II\rbrack&4076.35&$7.2\pm1.5$&$281.9\pm0.7$&$10.50\pm0.55$&$8.3\pm1.1$&$292.1\pm0.7$&$12.56\pm0.59$&1.90\\
H$\delta$&4101.73&$26.5\pm2.1$&$281.4\pm1.0$&$25.95\pm0.45$&$26.1\pm2.3$&$291.0\pm1.0$&$26.91\pm0.48$&1.89\\
H$\gamma$&4340.46&$48.4\pm1.9$&$281.2\pm0.8$&$24.56\pm0.28$&$48.2\pm2.2$&$291.3\pm0.8$&$26.69\pm0.25$&1.84\\
\lbrack O III\rbrack&4363.21&$11.2\pm1.2$&$280.0\pm1.1$&$21.19\pm0.71$&$8.6\pm1.2$&$292.8\pm1.2$&$20.20\pm1.19$&1.84\\
He II &4685.7&-&-&-&-&-&-&1.75\\
H$\beta$&4861.32&$100.0$&$282.0\pm0.5$&$25.68\pm0.12$&$105.9\pm2.1$&$290.9\pm0.4$&$26.69\pm0.07$&1.71\\
\lbrack O III\rbrack&4958.91&$58.5\pm1.9$&$281.7\pm0.5$&$17.82\pm0.17$&$41.0\pm1.3$&$294.4\pm0.6$&$21.92\pm0.25$&1.68\\
\lbrack O III\rbrack&5006.84&$176.1\pm4.6$&$281.7\pm0.4$&$17.94\pm0.09$&$124.7\pm3.5$&$294.8\pm0.5$&$21.95\pm0.12$&1.67\\
\lbrack O I\rbrack&5577.34&$0.5\pm0.2$&$282.2\pm1.3$&$13.29\pm1.17$&$0.5\pm0.2$&$289.4\pm1.6$&$17.75\pm1.38$&1.57\\
\lbrack N 
II\rbrack&5754.59&$21.3\pm0.6$&$282.0\pm0.3$&$13.84\pm0.08$&$20.5\pm0.6$&$292.4\pm0.4$&$14.87\pm0.16$&1.54\\
He I&5875.63&$20.6\pm0.8$&$281.2\pm0.5$&$16.51\pm0.14$&$19.3\pm0.8$&$290.7\pm0.5$&$17.77\pm0.18$&1.53\\
\lbrack O I\rbrack&6300.30&$52.2\pm1.4$&$281.5\pm0.3$&$11.39\pm0.07$&$52.3\pm1.1$&$291.9\pm0.3$&$13.20\pm0.08$&1.47\\
\lbrack S III\rbrack&6312.06&$1.7\pm0.3$&$281.2\pm1.2$&$14.96\pm1.13$&$1.1\pm0.3$&$293.4\pm1.5$&$17.47\pm1.59$&1.47\\
\lbrack O I\rbrack&6363.78&$18.0\pm0.5$&$281.3\pm0.3$&$11.29\pm0.05$&$18.5\pm0.5$&$292.2\pm0.3$&$13.69\pm0.09$&1.47\\
\lbrack N II\rbrack&6548.05&$394.7\pm8.6$&$282.3\pm0.2$&$12.84\pm0.04$&$397.5\pm7.3$&$292.7\pm0.2$&$14.18\pm0.04$&1.45\\
H$\alpha$&6562.80&$395.1\pm6.5$&$281.7\pm0.3$&$27.15\pm0.04$&$380.7\pm6.2$&$290.8\pm0.3$&$26.74\pm0.04$&1.45\\
\lbrack N II\rbrack&6583.45&$1206.3\pm24.2$&$280.8\pm0.2$&$12.79\pm0.04$&$1216.9\pm20.2$&$291.5\pm0.2$&$14.18\pm0.03$&1.45\\
He I &6678.15&$5.7\pm0.4$&$282.0\pm0.6$&$15.91\pm0.24$&$5.6\pm0.3$&$292.1\pm0.7$&$17.98\pm0.44$&1.44\\
\lbrack S II\rbrack&6716.44&$72.8\pm2.1$&$281.9\pm0.3$&$11.82\pm0.07$&$76.2\pm1.9$&$292.9\pm0.3$&$13.24\pm0.06$&1.43\\
\lbrack S II\rbrack&6730.82&$121.1\pm2.9$&$281.9\pm0.3$&$11.40\pm0.05$&$128.3\pm2.7$&$292.8\pm0.2$&$13.02\pm0.05$&1.43\\
\lbrack Ar III\rbrack&7135.79&$6.1\pm0.4$&$280.3\pm0.5$&$14.86\pm0.26$&$4.6\pm0.4$&$292.2\pm0.7$&$16.39\pm0.62$&1.39\\
\lbrack Fe II\rbrack&7155.16&$5.1\pm0.2$&$281.7\pm0.4$&$11.03\pm0.11$&$5.3\pm0.2$&$293.8\pm0.3$&$10.63\pm0.23$&1.39\\
\lbrack O II\rbrack&7319$^{\mathrm{d}}$&$13.2\pm0.8$&-&-&$13.2\pm0.9$&-&-&1.37\\
\lbrack Ca II\rbrack&7323.89&$11.5\pm0.4$&$280.6\pm0.3$&$10.82\pm0.08$&$11.2\pm0.5$&$292.3\pm0.4$&$11.06\pm0.15$&1.37\\
\lbrack O II\rbrack&7330$^{\mathrm{e}}$&$10.7\pm0.7$&-&-&$11.1\pm0.7$&-&-&1.37\\
\lbrack S III\rbrack&9068.6&$14.0\pm0.9$&-&$10.65\pm0.24$&$8.4\pm0.4$&-&$15.03\pm0.25$&1.24\\
\lbrack S III\rbrack&9530.6+31.1&$30.7\pm1.6$&-&$14.60\pm0.24$&$18.1\pm0.8$&-&$17.58\pm0.22$&1.22\\
\hline
\label{tab:narrowlines2}
\end{tabular}
\begin{list}{}{}
 \item[$^{\mathrm{a}}$] Fluxes are relative to H$\beta$: ``100" corresponds
 to $(29.3\pm 0.5) \times 10^{-16}\ergsm$.
 \item[$^{\mathrm{b}}$] $E(B-V)=0.16$, with $E(B-V)=0.10$ for LMC and $E(B-V)=0.06$ for the Milky Way.
 \item[$^{\mathrm{c}}$] The recession velocity of the peak flux.
\item[$^{\mathrm{d}}$] Applies to both lines at $7318.92\ang$ and $7319.99\ang$. 
\item[$^{\mathrm{e}}$] Applies to both lines at $7329.67\ang$ and $7330.73\ang$.
\end{list}
\end{table*}

\begin{table*}
\caption{Emission lines from the unshocked gas of the ER at Epoch 3.}
\begin{tabular}{llccccccc}
\hline
\hline
&&&North&&&South&&\\
 & Rest wavel.& &$V_{\rm peak}^{\mathrm{c}}$&$V_{\rm FWHM}$&&$V_{\rm peak}$&$V_{\rm FWHM}$&Extinction\\
Emission&$\ang$&Relative flux$^{\mathrm{a}}$&$\kms$&$\kms$&Relative flux$^{\mathrm{a}}$&$\kms$&$\kms$&correction$^{\mathrm{b}}$\\
\hline
\lbrack Ne V\rbrack&3425.86&$17.2\pm1.7$&$281.7\pm1.2$&$23.45\pm0.58$&$13.8\pm1.6$&$294.4\pm1.0$&$19.55\pm0.57$&2.06\\
\lbrack O II\rbrack&3726.03&$109.2\pm4.1$&$283.0\pm0.4$&$13.82\pm0.13$&$101.1\pm4.0$&$293.3\pm0.5$&$14.77\pm0.23$&1.98\\
\lbrack O II\rbrack&3728.82&$66.4\pm3.7$&$280.1\pm0.5$&$14.19\pm0.20$&$63.8\pm3.9$&$290.6\pm0.6$&$14.87\pm0.37$&1.98\\
\lbrack Ne III\rbrack&3868.75&$44.0\pm1.8$&$281.3\pm0.5$&$15.72\pm0.18$&$24.7\pm1.8$&$292.1\pm0.7$&$17.64\pm0.28$&1.94\\
\lbrack S II\rbrack&4068.60&$17.0\pm0.9$&$281.3\pm0.4$&$11.07\pm0.15$&$15.2\pm0.8$&$291.3\pm0.6$&$13.77\pm0.22$&1.90\\
\lbrack S II\rbrack&4076.35&$5.4\pm0.6$&$281.7\pm0.6$&$11.46\pm0.30$&$5.1\pm0.6$&$292.2\pm0.9$&$13.60\pm0.44$&1.90\\
H$\delta$&4101.73&$20.3\pm0.8$&$281.7\pm1.0$&$28.43\pm0.36$&$15.6\pm0.6$&$290.9\pm0.8$&$24.72\pm0.29$&1.89\\
H$\gamma$&4340.46&$44.6\pm1.7$&$281.1\pm1.0$&$25.86\pm0.26$&$39.4\pm1.6$&$291.0\pm0.8$&$26.52\pm0.24$&1.84\\
\lbrack O III\rbrack&4363.21&$10.6\pm0.7$&$279.3\pm0.7$&$17.55\pm0.31$&$6.7\pm0.6$&$290.8\pm1.0$&$23.97\pm0.64$&1.84\\
He II &4685.7&$10.1\pm1.4$&$283.0\pm1.2$&$21.56\pm1.27$&$7.1\pm0.7$&$295.6\pm1.6$&$30.39\pm1.41$&1.75\\
H$\beta$&4861.32&$100.0$&$280.9\pm0.5$&$26.04\pm0.11$&$97.9\pm2.6$&$290.6\pm0.6$&$27.36\pm0.16$&1.71\\
\lbrack O III\rbrack&4958.91&$81.3\pm2.1$&$281.0\pm0.4$&$15.69\pm0.11$&$50.1\pm1.3$&$291.8\pm0.5$&$22.38\pm0.14$&1.68\\
\lbrack O III\rbrack&5006.84&$247.2\pm5.1$&$281.1\pm0.3$&$15.88\pm0.06$&$154.9\pm3.5$&$292.5\pm0.5$&$22.85\pm0.10$&1.67\\
\lbrack O I\rbrack&5577.34&$0.5\pm0.2$&$283.4\pm1.4$&$14.66\pm1.17$&$0.5\pm0.2$&$289.5\pm1.4$&$15.19\pm1.23$&1.57\\
\lbrack N 
II\rbrack&5754.59&$20.6\pm0.6$&$281.5\pm0.3$&$13.69\pm0.09$&$19.3\pm0.5$&$292.0\pm0.4$&$16.21\pm0.16$&1.54\\
He I&5875.63&$17.3\pm0.6$&$280.8\pm0.5$&$16.85\pm0.15$&$16.9\pm0.6$&$290.3\pm0.5$&$18.63\pm0.16$&1.53\\
\lbrack O I\rbrack&6300.30&$48.0\pm1.9$&$281.4\pm0.4$&$11.67\pm0.12$&$47.7\pm1.4$&$291.2\pm0.4$&$14.12\pm0.16$&1.47\\
\lbrack S III\rbrack&6312.06&$2.6\pm0.3$&$280.6\pm0.7$&$12.78\pm0.47$&$1.3\pm0.2$&$291.9\pm1.3$&$18.49\pm0.85$&1.47\\
\lbrack O I\rbrack&6363.78&$16.1\pm0.6$&$281.1\pm0.3$&$11.40\pm0.09$&$16.7\pm0.4$&$291.6\pm0.4$&$14.50\pm0.12$&1.47\\
\lbrack N II\rbrack&6548.05&$359.9\pm7.3$&$281.9\pm0.3$&$12.95\pm0.05$&$352.7\pm6.9$&$292.5\pm0.3$&$14.88\pm0.05$&1.45\\
H$\alpha$&6562.80&$419.2\pm6.9$&$280.9\pm0.4$&$28.69\pm0.07$&$369.8\pm5.7$&$290.5\pm0.3$&$27.82\pm0.05$&1.45\\
\lbrack N II\rbrack&6583.45&$1112.7\pm18.7$&$280.7\pm0.2$&$12.74\pm0.03$&$1096.4\pm18.2$&$291.2\pm0.2$&$14.99\pm0.04$&1.45\\
He I &6678.15&$5.5\pm0.6$&$281.1\pm0.7$&$16.07\pm0.35$&$4.7\pm0.3$&$291.6\pm0.7$&$18.12\pm0.32$&1.44\\
\lbrack S II\rbrack&6716.44&$62.2\pm1.5$&$281.9\pm0.3$&$11.61\pm0.06$&$69.6\pm1.8$&$292.9\pm0.3$&$13.75\pm0.07$&1.43\\
\lbrack S II\rbrack&6730.82&$103.5\pm2.4$&$281.7\pm0.3$&$11.46\pm0.06$&$115.5\pm3.3$&$292.7\pm0.3$&$13.82\pm0.08$&1.43\\
\lbrack Ar III\rbrack&7135.79&$8.3\pm0.4$&$279.9\pm0.4$&$12.91\pm0.15$&$5.4\pm0.4$&$290.1\pm0.7$&$18.11\pm0.29$&1.39\\
\lbrack Fe II\rbrack&7155.16&$4.8\pm0.3$&$281.3\pm0.4$&$9.91\pm0.13$&$4.9\pm0.2$&$293.2\pm0.4$&$13.73\pm0.28$&1.39\\
\lbrack O II\rbrack&7319$^{\mathrm{d}}$&$13.1\pm1.0$&-&-&$12.3\pm0.7$&-&-&1.37\\
\lbrack Ca II\rbrack&7323.89&$7.2\pm0.3$&$279.9\pm0.4$&$9.77\pm0.13$&$9.8\pm0.3$&$291.9\pm0.4$&$13.39\pm0.21$&1.37\\
\lbrack O II\rbrack&7330$^{\mathrm{e}}$&$11.4\pm0.8$&-&-&$9.5\pm0.8$&-&-&1.37\\
\lbrack S III\rbrack&9068.6&$21.5\pm0.7$&-&$12.76\pm0.10$&$11.2\pm0.4$&-&$18.90\pm0.17$&1.24\\
\lbrack S III\rbrack&9530.6+31.1&$55.2\pm2.0$&-&$12.84\pm0.12$&$27.8\pm0.9$&-&$17.58\pm0.22$&1.22\\
\hline
\label{tab:narrowlines3}
\end{tabular}
\begin{list}{}{}
 \item[$^{\mathrm{a}}$] Fluxes are relative to H$\beta$: ``100" corresponds
 to $(31.3\pm 0.5) \times 10^{-16}\ergsm$.
 \item[$^{\mathrm{b}}$] $E(B-V)=0.16$, with $E(B-V)=0.10$ for LMC and $E(B-V)=0.06$ for the Milky Way.
 \item[$^{\mathrm{c}}$] The recession velocity of the peak flux.
\item[$^{\mathrm{d}}$] Applies to both lines at $7318.92\ang$ and $7319.99\ang$. 
\item[$^{\mathrm{e}}$] Applies to both lines at $7329.67\ang$ and $7330.73\ang$. 
\end{list}
\end{table*}

\begin{table*}
\caption{Emission lines from the unshocked gas of the ER at Epoch 4.}
\begin{tabular}{llccccccc}
\hline
\hline
&&&North&&&South&&\\
 & Rest wavel.& &$V_{\rm peak}^{\mathrm{c}}$&$V_{\rm FWHM}$&&$V_{\rm peak}$&$V_{\rm FWHM}$&Extinction\\
Emission&$\ang$&Relative flux$^{\mathrm{a}}$&$\kms$&$\kms$&Relative flux$^{\mathrm{a}}$&$\kms$&$\kms$&correction$^{\mathrm{b}}$\\
\hline
\lbrack Ne V\rbrack&3425.86&$18.3\pm2.8$&$281.7\pm1.2$&$20.31\pm0.76$&$15.1\pm1.8$&$293.4\pm1.6$&$20.41\pm0.71$&2.06\\
\lbrack O II\rbrack&3726.03&$98.8\pm4.1$&$283.6\pm0.4$&$13.65\pm0.11$&$99.7\pm4.5$&$293.3\pm0.5$&$14.79\pm0.18$&1.98\\
\lbrack O II\rbrack&3728.82&$61.6\pm4.2$&$280.7\pm0.5$&$14.06\pm0.22$&$63.9\pm4.6$&$290.3\pm0.6$&$15.90\pm0.21$&1.98\\
\lbrack Ne III\rbrack&3868.75&$59.1\pm3.2$&$282.2\pm0.5$&$15.40\pm0.20$&$49.0\pm3.6$&$292.6\pm0.6$&$17.57\pm0.23$&1.94\\
\lbrack S II\rbrack&4068.60&$11.2\pm2.2$&$281.9\pm0.9$&$12.45\pm0.65$&$11.4\pm1.5$&$292.9\pm0.8$&$12.29\pm0.66$&1.90\\
\lbrack S II\rbrack&4076.35&$4.2\pm0.7$&$283.8\pm1.3$&$17.64\pm1.02$&$3.0\pm0.9$&$292.5\pm1.7$&$14.54\pm1.61$&1.90\\
H$\delta$&4101.73&$20.7\pm1.4$&$281.6\pm1.5$&$31.70\pm0.71$&$18.4\pm1.2$&$292.4\pm1.4$&$29.94\pm0.68$&1.89\\
H$\gamma$&4340.46&$41.7\pm2.6$&$282.2\pm1.0$&$28.73\pm0.45$&$39.2\pm3.3$&$292.2\pm1.1$&$27.07\pm0.53$&1.84\\
\lbrack O III\rbrack&4363.21&$10.4\pm1.0$&$280.1\pm0.8$&$16.70\pm0.52$&$8.1\pm1.0$&$290.8\pm1.0$&$17.33\pm0.64$&1.84\\
He II &4685.7&$10.9\pm1.4$&$285.2\pm1.8$&$29.07\pm1.34$&$7.2\pm1.2$&$297.2\pm1.8$&$24.77\pm2.20$&1.75\\
H$\beta$&4861.32&$100.0$&$281.6\pm0.9$&$32.06\pm0.25$&$100.7\pm3.9$&$292.2\pm0.7$&$28.68\pm0.18$&1.71\\
\lbrack O III\rbrack&4958.91&$94.6\pm3.4$&$282.0\pm0.3$&$16.02\pm0.07$&$75.8\pm2.9$&$293.1\pm0.5$&$19.87\pm0.14$&1.68\\
\lbrack O III\rbrack&5006.84&$278.6\pm8.9$&$282.2\pm0.3$&$16.28\pm0.04$&$229.6\pm7.2$&$292.9\pm0.4$&$20.09\pm0.05$&1.67\\
\lbrack O I\rbrack&5577.34&$0.7\pm0.2$&$282.7\pm1.5$&$15.83\pm1.30$&$0.6\pm0.2$&$289.8\pm1.6$&$15.03\pm1.48$&1.57\\
\lbrack N 
II\rbrack&5754.59&$18.0\pm0.9$&$282.3\pm0.4$&$14.36\pm0.15$&$17.2\pm0.8$&$292.0\pm0.5$&$15.84\pm0.21$&1.54\\
He I&5875.63&$11.7\pm0.8$&$281.4\pm0.7$&$17.74\pm0.29$&$12.2\pm0.7$&$291.7\pm0.7$&$18.49\pm0.27$&1.53\\
\lbrack O I\rbrack&6300.30&$34.1\pm3.5$&$282.4\pm0.7$&$12.1\pm0.4$&$37.4\pm2.3$&$292.4\pm0.6$&$13.09\pm0.27$&1.47\\
\lbrack S III\rbrack&6312.06&$2.7\pm0.4$&$281.8\pm1.1$&$15.18\pm1.20$&$2.8\pm0.4$&$291.3\pm1.0$&$17.67\pm0.83$&1.47\\
\lbrack O I\rbrack&6363.78&$11.9\pm0.6$&$282.1\pm0.4$&$12.16\pm0.10$&$13.6\pm0.5$&$291.9\pm0.4$&$14.11\pm0.10$&1.47\\
\lbrack N II\rbrack&6548.05&$263.3\pm8.2$&$282.8\pm0.2$&$13.42\pm0.03$&$281.1\pm8.6$&$292.7\pm0.2$&$14.64\pm0.03$&1.45\\
H$\alpha$&6562.80&$394.2\pm12.1$&$281.6\pm0.5$&$31.64\pm0.07$&$380.6\pm11.7$&$291.0\pm0.4$&$29.49\pm0.07$&1.45\\
\lbrack N II\rbrack&6583.45&$812.7\pm24.7$&$281.5\pm0.2$&$13.28\pm0.02$&$866.5\pm26.3$&$291.5\pm0.2$&$14.63\pm0.03$&1.45\\
He I &6678.15&$3.9\pm0.5$&$282.7\pm1.0$&$19.86\pm0.47$&$3.4\pm0.3$&$292.0\pm1.5$&$17.40\pm0.38$&1.44\\
\lbrack S II\rbrack&6716.44&$39.3\pm1.4$&$282.9\pm0.3$&$12.05\pm0.05$&$48.5\pm1.8$&$293.2\pm0.3$&$13.26\pm0.07$&1.43\\
\lbrack S II\rbrack&6730.82&$63.1\pm2.2$&$282.8\pm0.2$&$11.88\pm0.04$&$74.0\pm2.5$&$293.4\pm0.4$&$13.49\pm0.09$&1.43\\
\lbrack Ar III\rbrack&7135.79&$9.8\pm0.5$&$280.5\pm0.4$&$14.02\pm0.12$&$7.9\pm0.6$&$291.2\pm0.7$&$16.92\pm0.37$&1.39\\
\lbrack Fe II\rbrack&7155.16&$3.3\pm0.4$&$281.6\pm0.6$&$11.44\pm0.29$&$3.7\pm0.4$&$294.1\pm0.5$&$10.23\pm0.29$&1.39\\
\lbrack O II\rbrack&7319$^{\mathrm{d}}$&$9.4\pm1.6$&-&-&$10.8\pm1.5$&-&-&1.37\\
\lbrack Ca II\rbrack&7323.89&$2.1\pm0.2$&$280.6\pm0.5$&$9.66\pm0.22$&$3.7\pm0.3$&$293.1\pm0.5$&$10.48\pm0.23$&1.37\\
\lbrack O II\rbrack&7330$^{\mathrm{e}}$&$12.6\pm1.7$&-&-&$9.0\pm0.9$&-&-&1.37\\
\lbrack S III\rbrack&9068.6&-&-&-&-&-&-&1.24\\
\lbrack S III\rbrack&9530.6+31.1&-&-&-&-&-&-&1.22\\
\hline
\label{tab:narrowlines4}
\end{tabular}
\begin{list}{}{}
 \item[$^{\mathrm{a}}$] Fluxes are relative to H$\beta$: ``100" corresponds
 to $(32.6\pm 1.0) \times 10^{-16}\ergsm$.
 \item[$^{\mathrm{b}}$] $E(B-V)=0.16$, with $E(B-V)=0.10$ for LMC and $E(B-V)=0.06$ for the Milky Way.
 \item[$^{\mathrm{c}}$] The recession velocity of the peak flux.
\item[$^{\mathrm{d}}$] Applies to both lines at $7318.92\ang$ and $7319.99\ang$. 
\item[$^{\mathrm{e}}$] Applies to both lines at $7329.67\ang$ and $7330.73\ang$. 
\end{list}
\end{table*}

\begin{table*}
\caption{Emission lines from the unshocked gas of the ER at Epoch 5.}
\begin{tabular}{llccccccc}
\hline
\hline
&&&North&&&South&&\\
 & Rest wavel.& &$V_{\rm peak}^{\mathrm{c}}$&$V_{\rm FWHM}$&&$V_{\rm peak}$&$V_{\rm FWHM}$&Extinction\\
Emission&$\ang$&Relative flux$^{\mathrm{a}}$&$\kms$&$\kms$&Relative flux$^{\mathrm{a}}$&$\kms$&$\kms$&correction$^{\mathrm{b}}$\\
\hline
\lbrack Ne V\rbrack&3425.86&$18.5\pm2.2$&$282.6\pm1.4$&$21.72\pm0.98$&$18.5\pm2.2$&$291.7\pm1.5$&$23.03\pm1.02$&2.06\\
\lbrack O II\rbrack&3726.03&$94.6\pm6.4$&$281.9\pm0.6$&$14.13\pm0.21$&$108.4\pm4.8$&$291.5\pm0.6$&$16.33\pm0.19$&1.98\\
\lbrack O II\rbrack&3728.82&$59.5\pm4.8$&$278.9\pm0.6$&$13.42\pm0.30$&$68.0\pm5.2$&$289.7\pm0.9$&$16.86\pm0.39$&1.98\\
\lbrack Ne III\rbrack&3868.75&$61.4\pm3.1$&$282.4\pm0.5$&$15.03\pm0.17$&$50.6\pm2.3$&$291.9\pm0.5$&$17.48\pm0.14$&1.94\\
\lbrack S II\rbrack&4068.60&$10.7\pm1.2$&$281.7\pm0.6$&$12.00\pm0.41$&$11.3\pm1.1$&$292.1\pm0.7$&$12.45\pm0.54$&1.90\\
\lbrack S II\rbrack&4076.35&$3.2\pm0.5$&$282.0\pm1.0$&$13.58\pm0.74$&$3.7\pm0.5$&$291.2\pm1.0$&$14.52\pm0.71$&1.90\\
H$\delta$&4101.73&$21.1\pm1.6$&$282.7\pm1.1$&$28.29\pm0.64$&$20.5\pm1.4$&$292.7\pm1.2$&$29.17\pm0.78$&1.89\\
H$\gamma$&4340.46&$40.0\pm1.9$&$280.8\pm0.9$&$30.56\pm0.35$&$42.4\pm1.9$&$291.4\pm0.9$&$30.13\pm0.30$&1.84\\
\lbrack O III\rbrack&4363.21&$9.8\pm0.7$&$280.1\pm0.6$&$15.91\pm0.31$&$9.7\pm0.8$&$290.5\pm0.4$&$19.46\pm0.34$&1.84\\
He II &4685.7&$9.5\pm1.1$&$284.2\pm1.3$&$26.66\pm0.86$&$9.4\pm0.8$&$294.6\pm1.3$&$29.04\pm0.80$&1.75\\
H$\beta$&4861.32&$100.0$&$281.1\pm0.8$&$30.65\pm0.25$&$122.2\pm4.5$&$290.9\pm0.7$&$30.31\pm0.20$&1.71\\
\lbrack O III\rbrack&4958.91&$100.6\pm4.9$&$281.9\pm0.4$&$15.42\pm0.12$&$97.3\pm4.2$&$290.1\pm0.5$&$19.26\pm0.14$&1.68\\
\lbrack O III\rbrack&5006.84&$303.6\pm12.6$&$282.0\pm0.4$&$15.53\pm0.11$&$287.9\pm9.2$&$290.2\pm0.4$&$19.40\pm0.08$&1.67\\
\lbrack O I\rbrack&5577.34&$0.6\pm0.3$&$282.1\pm2.0$&$13.58\pm1.94$&$0.7\pm0.4$&$289.4\pm2.7$&$16.10\pm2.64$&1.57\\
\lbrack N 
II\rbrack&5754.59&$18.8\pm2.0$&$281.8\pm0.6$&$12.89\pm0.29$&$21.3\pm1.4$&$291.4\pm0.7$&$16.56\pm0.38$&1.54\\
He I&5875.63&$12.0\pm1.9$&$281.3\pm1.1$&$17.32\pm0.65$&$14.4\pm1.5$&$289.5\pm1.0$&$19.52\pm0.52$&1.53\\
\lbrack O I\rbrack&6300.30&$33.8\pm4.8$&$281.5\pm0.8$&$11.36\pm0.42$&$43.1\pm2.7$&$289.8\pm0.6$&$14.02\pm0.33$&1.47\\
\lbrack S III\rbrack&6312.06&$3.6\pm0.9$&$281.8\pm0.8$&$11.00\pm0.65$&$3.1\pm0.6$&$289.2\pm1.4$&$17.21\pm0.82$&1.47\\
\lbrack O I\rbrack&6363.78&$13.0\pm1.0$&$281.4\pm0.5$&$11.76\pm0.18$&$16.1\pm0.9$&$290.2\pm0.5$&$14.24\pm0.22$&1.47\\
\lbrack N II\rbrack&6548.05&$271.2\pm9.4$&$282.2\pm0.3$&$12.78\pm0.06$&$326.3\pm10.5$&$291.8\pm0.3$&$15.74\pm0.05$&1.45\\
H$\alpha$&6562.80&$418.7\pm13.3$&$280.5\pm0.6$&$31.26\pm0.13$&$500.5\pm15.7$&$290.0\pm0.6$&$31.09\pm0.11$&1.45\\
\lbrack N II\rbrack&6583.45&$838.3\pm26.8$&$280.8\pm0.3$&$12.76\pm0.04$&$992.3\pm31.2$&$290.5\pm0.3$&$15.48\pm0.05$&1.45\\
He I &6678.15&$4.2\pm1.1$&$281.3\pm1.6$&$19.31\pm1.27$&$4.2\pm0.8$&$287.4\pm1.5$&$18.68\pm0.83$&1.44\\
\lbrack S II\rbrack&6716.44&$38.4\pm1.8$&$282.2\pm0.3$&$11.48\pm0.10$&$52.0\pm2.1$&$292.3\pm0.4$&$14.39\pm0.14$&1.43\\
\lbrack S II\rbrack&6730.82&$59.9\pm2.9$&$282.2\pm0.3$&$10.96\pm0.10$&$80.1\pm3.0$&$292.2\pm0.4$&$14.09\pm0.12$&1.43\\
\lbrack Ar III\rbrack&7135.79&$9.8\pm0.5$&$280.8\pm0.4$&$13.34\pm0.14$&$10.1\pm0.6$&$289.8\pm0.5$&$16.39\pm0.15$&1.39\\
\lbrack Fe II\rbrack&7155.16&$3.2\pm0.3$&$281.4\pm0.4$&$9.67\pm0.14$&$3.8\pm0.3$&$293.3\pm0.5$&$11.04\pm0.27$&1.39\\
\lbrack O II\rbrack&7319$^{\mathrm{d}}$&$9.0\pm1.5$&-&-&$12.3\pm1.1$&-&-&1.37\\
\lbrack Ca II\rbrack&7323.89&$1.4\pm0.2$&$279.9\pm0.6$&$10.01\pm0.40$&$2.8\pm0.2$&$292.7\pm0.4$&$9.37\pm0.13$&1.37\\
\lbrack O II\rbrack&7330$^{\mathrm{e}}$&$12.2\pm0.9$&-&-&$10.1\pm0.6$&-&-&1.37\\
\lbrack S III\rbrack&9068.6&-&-&-&-&-&-&1.24\\
\lbrack S III\rbrack&9530.6+31.1&-&-&-&-&-&-&1.22\\
\hline
\label{tab:narrowlines5}
\end{tabular}
\begin{list}{}{}
 \item[$^{\mathrm{a}}$] Fluxes are relative to H$\beta$: ``100" corresponds
 to $(31.8\pm 1.0) \times 10^{-16}\ergsm$.
 \item[$^{\mathrm{b}}$] $E(B-V)=0.16$, with $E(B-V)=0.10$ for LMC and $E(B-V)=0.06$ for the Milky Way.
 \item[$^{\mathrm{c}}$] The recession velocity of the peak flux.
\item[$^{\mathrm{d}}$] Applies to both lines at $7318.92\ang$ and $7319.99\ang$. 
\item[$^{\mathrm{e}}$] Applies to both lines at $7329.67\ang$ and $7330.73\ang$. 
\end{list}
\end{table*}

\begin{table*}
\caption{Emission lines from the unshocked gas of the ER at Epoch 6.}
\begin{tabular}{llccccccc}
\hline
\hline
&&&North&&&South&&\\
 & Rest wavel.& &$V_{\rm peak}^{\mathrm{c}}$&$V_{\rm FWHM}$&&$V_{\rm peak}$&$V_{\rm FWHM}$&Extinction\\
Emission&$\ang$&Relative flux$^{\mathrm{a}}$&$\kms$&$\kms$&Relative flux$^{\mathrm{a}}$&$\kms$&$\kms$&correction$^{\mathrm{b}}$\\
\hline
\lbrack Ne V\rbrack&3425.86&$20.9\pm2.1$&$282.9\pm1.4$&$23.70\pm0.88$&$20.1\pm1.5$&$294.2\pm1.0$&$20.37\pm0.53$&2.06\\
\lbrack O II\rbrack&3726.03&$107.8\pm5.2$&$283.6\pm0.4$&$14.51\pm0.13$&$119.9\pm5.8$&$293.6\pm0.4$&$14.06\pm0.13$&1.98\\
\lbrack O II\rbrack&3728.82&$67.0\pm4.0$&$280.7\pm0.5$&$14.52\pm0.18$&$74.9\pm3.7$&$290.9\pm0.4$&$14.17\pm0.14$&1.98\\
\lbrack Ne III\rbrack&3868.75&$69.5\pm3.4$&$282.1\pm0.4$&$14.51\pm0.17$&$66.4\pm4.2$&$291.4\pm0.5$&$16.02\pm0.19$&1.94\\
\lbrack S II\rbrack&4068.60&$8.6\pm0.8$&$281.8\pm0.7$&$12.54\pm0.39$&$10.1\pm0.9$&$291.3\pm0.6$&$12.49\pm0.33$&1.90\\
\lbrack S II\rbrack&4076.35&$3.4\pm0.6$&$282.2\pm1.0$&$12.66\pm0.77$&$3.8\pm0.5$&$291.1\pm0.8$&$13.20\pm0.57$&1.90\\
H$\delta$&4101.73&$17.0\pm1.9$&$282.4\pm1.8$&$30.15\pm1.60$&$22.0\pm1.9$&$290.5\pm1.4$&$28.56\pm0.70$&1.89\\
H$\gamma$&4340.46&$36.8\pm2.2$&$280.3\pm1.2$&$30.67\pm0.48$&$52.3\pm3.4$&$291.0\pm0.9$&$31.39\pm0.34$&1.84\\
\lbrack O III\rbrack&4363.21&$13.0\pm1.0$&$279.9\pm0.9$&$20.40\pm0.41$&$12.8\pm1.0$&$290.0\pm0.8$&$19.60\pm0.36$&1.84\\
He II &4685.7&$10.9\pm1.5$&$283.9\pm1.5$&$28.01\pm0.99$&$13.3\pm1.3$&$294.7\pm1.3$&$26.84\pm1.19$&1.75\\
H$\beta$&4861.32&100.0&$280.1\pm0.8$&$28.75\pm0.28$&$138.8\pm5.3$&$291.2\pm0.7$&$31.29\pm0.15$&1.71\\
\lbrack O III\rbrack&4958.91&$122.9\pm4.5$&$281.7\pm0.3$&$15.63\pm0.06$&$129.6\pm5.3$&$290.2\pm0.3$&$17.77\pm0.06$&1.68\\
\lbrack O III\rbrack&5006.84&$365.4\pm13.6$&$282.0\pm0.3$&$15.76\pm0.09$&$375.4\pm14.8$&$290.9\pm0.4$&$17.54\pm0.12$&1.67\\
\lbrack O I\rbrack&5577.34&$0.7\pm0.2$&$282.4\pm1.2$&$11.91\pm1.28$&$0.9\pm0.3$&$291.3\pm1.7$&$16.43\pm1.53$&1.57\\
\lbrack N 
II\rbrack&5754.59&$19.2\pm3.2$&$282.0\pm0.4$&$12.98\pm0.16$&$24.9\pm1.6$&$292.4\pm0.7$&$16.34\pm0.54$&1.54\\
He I&5875.63&$11.6\pm1.2$&$281.0\pm0.8$&$17.03\pm0.44$&$14.8\pm1.0$&$291.3\pm0.7$&$19.09\pm0.42$&1.53\\
\lbrack O I\rbrack&6300.30&$35.4\pm7.4$&$281.7\pm1.0$&$11.41\pm0.63$&$47.0\pm4.7$&$290.6\pm0.7$&$12.91\pm0.43$&1.47\\
\lbrack S III\rbrack&6312.06&$4.1\pm0.5$&$281.2\pm0.6$&$13.42\pm0.39$&$4.3\pm0.6$&$291.0\pm0.8$&$16.39\pm0.42$&1.47\\
\lbrack O I\rbrack&6363.78&$13.5\pm1.1$&$281.8\pm0.5$&$11.46\pm0.18$&$17.8\pm1.2$&$291.6\pm0.6$&$13.23\pm0.33$&1.47\\
\lbrack N II\rbrack&6548.05&$285.9\pm9.4$&$282.3\pm0.2$&$13.14\pm0.03$&$359.3\pm11.6$&$292.7\pm0.2$&$14.29\pm0.03$&1.45\\
H$\alpha$&6562.80&$424.2\pm15.0$&$280.8\pm0.6$&$30.42\pm0.16$&$621.4\pm20.9$&$291.0\pm0.5$&$31.68\pm0.10$&1.45\\
\lbrack N II\rbrack&6583.45&$880.5\pm28.8$&$281.0\pm0.2$&$12.97\pm0.03$&$1109.4\pm35.7$&$291.3\pm0.2$&$14.18\pm0.03$&1.45\\
He I &6678.15&$2.9\pm0.3$&$281.8\pm0.9$&$16.20\pm0.58$&$4.4\pm0.3$&$291.1\pm0.8$&$18.58\pm0.34$&1.44\\
\lbrack S II\rbrack&6716.44&$38.0\pm1.8$&$282.4\pm0.3$&$11.69\pm0.08$&$53.1\pm1.9$&$292.9\pm0.3$&$12.58\pm0.09$&1.43\\
\lbrack S II\rbrack&6730.82&$58.7\pm2.3$&$282.3\pm0.3$&$11.55\pm0.06$&$79.7\pm3.0$&$292.6\pm0.3$&$12.50\pm0.12$&1.43\\
\lbrack Ar III\rbrack&7135.79&$12.6\pm0.7$&$280.7\pm0.4$&$13.64\pm0.14$&$13.1\pm0.7$&$289.9\pm0.4$&$15.96\pm0.11$&1.39\\
\lbrack Fe II\rbrack&7155.16&$3.7\pm0.6$&$281.7\pm0.5$&$9.79\pm0.30$&$4.3\pm0.4$&$293.5\pm0.4$&$10.24\pm0.26$&1.39\\
\lbrack O II\rbrack&7319$^{\mathrm{d}}$&$9.7\pm0.7$&-&-&$15.7\pm1.6$&-&-&1.37\\
\lbrack Ca II\rbrack&7323.89&$0.7\pm0.1$&$278.9\pm0.5$&$7.51\pm0.37$&$2.6\pm0.4$&$292.9\pm0.7$&$8.99\pm0.52$&1.37\\
\lbrack O II\rbrack&7330$^{\mathrm{e}}$&$12.0\pm1.0$&-&-&$11.9\pm0.9$&-&-&1.37\\
\lbrack S III\rbrack&9068.6&-&-&-&-&-&-&1.24\\
\lbrack S III\rbrack&9530.6+31.1&-&-&-&-&-&-&1.22\\
\hline
\label{tab:narrowlines6}
\end{tabular}
\begin{list}{}{}
 \item[$^{\mathrm{a}}$] Fluxes are relative to H$\beta$: ``100" corresponds
 to $(23.9\pm 0.8) \times 10^{-16}\ergsm$.
 \item[$^{\mathrm{b}}$] $E(B-V)=0.16$, with $E(B-V)=0.10$ for LMC and $E(B-V)=0.06$ for the Milky Way.
 \item[$^{\mathrm{c}}$] The recession velocity of the peak flux.
\item[$^{\mathrm{d}}$] Applies to both lines at $7318.92\ang$ and $7319.99\ang$. 
\item[$^{\mathrm{e}}$] Applies to both lines at $7329.67\ang$ and $7330.73\ang$. 
\end{list}
\end{table*}

\begin{table*}
\caption{Emission lines from the unshocked gas of the ER at Epoch 7.}
\begin{tabular}{llccccccc}
\hline
\hline
&&&North&&&South&&\\
 & Rest wavel.& &$V_{\rm peak}^{\mathrm{c}}$&$V_{\rm FWHM}$&&$V_{\rm peak}$&$V_{\rm FWHM}$&Extinction\\
Emission&$\ang$&Relative flux$^{\mathrm{a}}$&$\kms$&$\kms$&Relative flux$^{\mathrm{a}}$&$\kms$&$\kms$&correction$^{\mathrm{b}}$\\
\hline
\lbrack Ne V\rbrack&3425.86&$23.2\pm3.4$&$280.3\pm1.1$&$18.71\pm0.71$&$23.7\pm3.0$&$291.8\pm1.6$&$20.71\pm0.68$&2.06\\
\lbrack O II\rbrack&3726.03&$105.1\pm6.0$&$282.7\pm0.5$&$13.91\pm0.17$&$123.7\pm5.5$&$292.4\pm0.5$&$14.38\pm0.18$&1.98\\
\lbrack O II\rbrack&3728.82&$66.9\pm6.6$&$279.7\pm0.6$&$13.44\pm0.32$&$76.8\pm4.7$&$289.5\pm0.6$&$14.95\pm0.38$&1.98\\
\lbrack Ne III\rbrack&3868.75&$64.9\pm6.7$&$283.5\pm0.9$&$16.73\pm0.41$&$63.9\pm4.6$&$292.9\pm0.5$&$15.44\pm0.18$&1.94\\
\lbrack S II\rbrack&4068.60&$6.9\pm1.0$&$284.3\pm0.9$&$12.09\pm0.64$&$6.9\pm1.6$&$292.8\pm0.9$&$8.91\pm0.78$&1.90\\
\lbrack S II\rbrack&4076.35&$3.1\pm0.8$&$283.5\pm1.4$&$14.01\pm1.30$&$3.4\pm0.6$&$292.7\pm1.0$&$13.70\pm0.79$&1.90\\
H$\delta$&4101.73&$16.7\pm1.2$&$281.5\pm1.6$&$34.37\pm0.85$&$21.0\pm2.3$&$291.7\pm2.1$&$33.58\pm1.38$&1.89\\
H$\gamma$&4340.46&$39.2\pm2.3$&$280.3\pm1.2$&$30.67\pm0.48$&$55.7\pm3.6$&$291.0\pm1.0$&$31.39\pm0.34$&1.84\\
\lbrack O III\rbrack&4363.21&$11.5\pm1.3$&$281.0\pm0.9$&$18.70\pm0.54$&$10.8\pm0.9$&$291.5\pm0.8$&$18.25\pm0.36$&1.84\\
He II &4685.7&$10.1\pm1.4$&$285.9\pm1.6$&$27.91\pm1.74$&$12.5\pm1.5$&$295.6\pm1.5$&$27.30\pm1.11$&1.75\\
H$\beta$&4861.32&100.0&$281.2\pm0.8$&$30.41\pm0.31$&$132.8\pm5.4$&$291.6\pm0.8$&$30.86\pm0.21$&1.71\\
\lbrack O III\rbrack&4958.91&$123.1\pm4.3$&$282.0\pm0.3$&$16.30\pm0.06$&$146.6\pm5.7$&$290.5\pm0.4$&$17.68\pm0.08$&1.68\\
\lbrack O III\rbrack&5006.84&$367.5\pm14.3$&$282.2\pm0.4$&$16.34\pm0.11$&$453.0\pm19.9$&$290.8\pm0.5$&$17.73\pm0.13$&1.67\\
\lbrack O I\rbrack&5577.34&$0.6\pm0.3$&$281.1\pm1.5$&$10.41\pm1.43$&$1.1\pm0.3$&$290.4\pm1.2$&$13.28\pm1.01$&1.57\\
\lbrack N 
II\rbrack&5754.59&$23.3\pm1.1$&$282.1\pm0.4$&$13.18\pm0.13$&$29.3\pm2.2$&$291.2\pm0.6$&$15.16\pm0.34$&1.54\\
He I&5875.63&$11.5\pm1.0$&$280.7\pm0.8$&$18.54\pm0.39$&$14.4\pm1.4$&$289.8\pm1.1$&$21.54\pm0.58$&1.53\\
\lbrack O I\rbrack&6300.30&$33.9\pm3.4$&$282.2\pm0.6$&$11.17\pm0.31$&$47.1\pm2.5$&$290.6\pm0.4$&$11.81\pm0.25$&1.47\\
\lbrack S III\rbrack&6312.06&$3.9\pm0.5$&$281.4\pm0.6$&$11.64\pm0.60$&$4.7\pm0.6$&$290.5\pm0.7$&$14.75\pm0.40$&1.47\\
\lbrack O I\rbrack&6363.78&$14.5\pm0.9$&$282.0\pm0.4$&$12.11\pm0.13$&$18.2\pm0.8$&$291.3\pm0.4$&$13.68\pm0.18$&1.47\\
\lbrack N II\rbrack&6548.05&$266.2\pm8.9$&$282.4\pm0.2$&$12.92\pm0.04$&$357.4\pm11.9$&$292.2\pm0.3$&$14.70\pm0.04$&1.45\\
H$\alpha$&6562.80&$455.5\pm15.6$&$280.7\pm0.7$&$32.17\pm0.16$&$627.0\pm20.7$&$290.3\pm0.5$&$31.50\pm0.10$&1.45\\
\lbrack N II\rbrack&6583.45&$816.8\pm26.7$&$281.2\pm0.2$&$12.85\pm0.03$&$1093.3\pm35.0$&$291.0\pm0.2$&$14.58\pm0.03$&1.45\\
He I &6678.15&$3.0\pm0.4$&$281.5\pm1.3$&$19.20\pm0.85$&$4.0\pm0.4$&$291.4\pm1.0$&$18.27\pm0.61$&1.44\\
\lbrack S II\rbrack&6716.44&$32.6\pm2.2$&$282.2\pm0.4$&$11.26\pm0.14$&$48.6\pm2.1$&$292.3\pm0.4$&$13.55\pm0.20$&1.43\\
\lbrack S II\rbrack&6730.82&$48.2\pm2.3$&$282.3\pm0.3$&$11.35\pm0.09$&$71.7\pm2.9$&$292.3\pm0.4$&$13.00\pm0.16$&1.43\\
\lbrack Ar III\rbrack&7135.79&$11.9\pm0.8$&$281.3\pm0.5$&$14.45\pm0.24$&$11.9\pm0.6$&$291.5\pm0.5$&$14.71\pm0.26$&1.39\\
\lbrack Fe II\rbrack&7155.16&$3.3\pm0.4$&$282.6\pm0.7$&$12.27\pm0.45$&$3.7\pm0.3$&$293.8\pm0.5$&$9.77\pm0.21$&1.39\\
\lbrack O II\rbrack&7319$^{\mathrm{d}}$&$8.8\pm1.4$&-&-&$13.5\pm1.5$&-&-&1.37\\
\lbrack Ca II\rbrack&7323.89&$0.9\pm0.3$&$281.2\pm2.2$&-&$2.0\pm0.5$&$293.0\pm0.7$&$7.94\pm0.63$&1.37\\
\lbrack O II\rbrack&7330$^{\mathrm{e}}$&$13.3\pm1.2$&-&-&$10.6\pm0.8$&-&-&1.37\\
\lbrack S III\rbrack&9068.6&-&-&-&-&-&-&1.24\\
\lbrack S III\rbrack&9530.6+31.1&-&-&-&-&-&-&1.22\\
\hline
\label{tab:narrowlines7}
\end{tabular}
\begin{list}{}{}
 \item[$^{\mathrm{a}}$] Fluxes are relative to H$\beta$: ``100" corresponds
 to $(22.4\pm 0.7) \times 10^{-16}\ergsm$.
 \item[$^{\mathrm{b}}$] $E(B-V)=0.16$, with $E(B-V)=0.10$ for LMC and $E(B-V)=0.06$ for the Milky Way.
 \item[$^{\mathrm{c}}$] The recession velocity of the peak flux.
\item[$^{\mathrm{d}}$] Applies to both lines at $7318.92\ang$ and $7319.99\ang$. 
\item[$^{\mathrm{e}}$] Applies to both lines at $7329.67\ang$ and $7330.73\ang$. 
\end{list}
\end{table*}

\begin{table*}
\caption{Fluxes of emission lines from the shocked gas from the northern ER$^{\mathrm{a}}$.}
\begin{tabular}{llccccccc}
\hline
\hline
Ion&$\lambda_{rest}$&Extinct.&5039 d&5703 d&6618 d&6826 d&7183 d&7559 d\\
\hline
\lbrack Ne V\rbrack&3425.86&2.06&-&$2.82\pm0.67$&$1.99\pm0.28$&$2.07\pm0.36$&$1.25\pm0.30$&$1.00\pm0.18$\\
\lbrack N I\rbrack&3466.50&2.05&$9.67\pm3.35$&$5.04\pm0.92$&$5.49\pm0.64$&$6.71\pm1.08$&$6.61\pm0.38$&$7.47\pm0.45$\\
\lbrack Ne III\rbrack&3868.75&1.94&$27.78\pm3.64$&$21.17\pm0.62$&$17.27\pm0.48$&$16.03\pm0.35$&$14.76\pm0.34$&$12.52\pm0.46$\\
\lbrack S II\rbrack&4068.60&1.90&$40.36\pm2.60$&$32.60\pm0.81$&$35.14\pm0.58$&$33.90\pm0.49$&$36.07\pm0.29$&$36.59\pm0.66$\\
\lbrack S II\rbrack&4076.35&1.90&$9.43\pm0.99$&$8.82\pm0.38$&$8.12\pm0.29$&$8.27\pm0.13$&$8.41\pm0.17$&$7.90\pm0.19$\\
H$\delta$&4101.73&1.89&$22.91\pm2.11$&$20.36\pm0.43$&$21.96\pm0.64$&$19.96\pm0.38$&$21.38\pm0.28$&$21.06\pm0.35$\\
H$\gamma$&4340.46&1.84&$50.16\pm2.30$&$42.72\pm0.77$&$41.53\pm0.55$&$40.69\pm0.58$&$41.26\pm0.57$&$40.83\pm0.38$\\
\lbrack O III\rbrack&4363.21&1.84&$7.49\pm1.07$&$5.99\pm0.35$&$3.69\pm0.32$&$3.11\pm0.30$&$2.50\pm0.19$&$2.25\pm0.20$\\
\lbrack Fe III\rbrack&4658.05&1.76&-&$1.87\pm0.20$&$1.48\pm0.07$&$1.32\pm0.05$&$1.36\pm0.08$&$1.23\pm0.10$\\
He II&4685.7&1.75&-&$7.07\pm0.30$&$7.25\pm0.22$&$6.64\pm0.11$&$6.58\pm0.14$&$5.88\pm0.18$\\
\lbrack O III\rbrack&4958.91&1.68&$8.87\pm0.55$&$8.37\pm0.28$&$4.55\pm0.11$&$4.83\pm0.11$&$3.12\pm0.05$&$2.83\pm0.06$\\
\lbrack O III\rbrack&5006.84&1.67&$27.44\pm0.66$&$27.02\pm0.39$&$14.77\pm0.13$&$14.77\pm0.28$&$12.21\pm0.13$&$10.81\pm0.13$\\
\lbrack N I\rbrack&5199$^{\mathrm{c}}$&1.63&$1.30\pm0.30$&$0.93\pm0.16$&$0.74\pm0.07$&$0.84\pm0.06$&$1.08\pm0.07$&$1.23\pm0.06$\\
\lbrack Fe XIV\rbrack&5302.86&1.61&$1.38\pm0.24$&$2.33\pm0.19$&$2.35\pm0.09$&$2.89\pm0.13$&$3.07\pm0.07$&$3.51\pm0.09$\\
\lbrack O I\rbrack&5577.34&1.57&$3.32\pm0.55$&$3.18\pm0.17$&$3.30\pm0.09$&$3.81\pm0.14$&$3.99\pm0.08$&$4.63\pm0.08$\\
\lbrack N II\rbrack&5754.59&1.54&$43.63\pm0.72$&$42.51\pm0.44$&$32.65\pm0.34$&$34.65\pm0.31$&$34.08\pm0.28$&$36.15\pm0.27$\\
He I&5875.63&1.53&$29.97\pm0.69$&$29.78\pm0.32$&$24.79\pm0.28$&$27.91\pm0.42$&$30.29\pm0.26$&$30.16\pm0.22$\\
\lbrack Fe VII\rbrack&6087.0&1.50&-&$0.44\pm0.06$&$0.26\pm0.03$&$0.29\pm0.05$&$0.26\pm0.03$&$0.17\pm0.03$\\
\lbrack O I\rbrack&6300.30&1.47&$49.44\pm0.95$&$45.29\pm0.41$&$42.68\pm0.67$&$50.10\pm0.68$&$55.33\pm0.50$&$59.61\pm0.48$\\
\lbrack S III\rbrack&6312.06&1.47&-&$2.33\pm0.15$&-&-&-&-\\
\lbrack O I\rbrack&6363.78&1.47&$16.81\pm0.68$&$15.49\pm0.43$&$14.48\pm0.41$&$17.08\pm0.49$&$18.70\pm0.24$&$20.27\pm0.15$\\
\lbrack Fe X\rbrack&6374.51&1.47&$1.05\pm0.21$&$1.87\pm0.14$&$1.39\pm0.08$&$1.80\pm0.13$&$1.59\pm0.11$&$1.51\pm0.10$\\
\lbrack N II\rbrack&6548.05&1.45&$16.39\pm0.53$&$14.35\pm0.26$&$9.38\pm0.17$&$9.66\pm0.33$&$10.60\pm0.12$&$11.09\pm0.14$\\
H$\alpha$&6562.80&1.45&$395.08\pm5.42$&$403.59\pm3.39$&$344.00\pm2.26$&$387.03\pm1.92$&$393.28\pm1.58$&$406.91\pm1.52$\\
\lbrack N II\rbrack&6583.45&1.45&$50.85\pm0.94$&$43.51\pm0.43$&$31.12\pm0.26$&$33.60\pm0.40$&$31.76\pm0.20$&$32.80\pm0.19$\\
He I&6678.15&1.44&$6.72\pm0.57$&$7.36\pm0.19$&$6.47\pm0.09$&$6.79\pm0.15$&$7.24\pm0.19$&$7.45\pm0.09$\\
\lbrack S II\rbrack&6716.44&1.43&$1.34\pm0.16$&$0.86\pm0.05$&$0.55\pm0.05$&$0.50\pm0.08$&$0.74\pm0.05$&$0.76\pm0.04$\\
\lbrack S II\rbrack&6730.82&1.43&$3.06\pm0.21$&$1.98\pm0.16$&$1.55\pm0.06$&$1.88\pm0.13$&$1.80\pm0.05$&$2.03\pm0.07$\\
\lbrack Ar V\rbrack&7005.67&1.41&-&$0.23\pm0.05$&$0.07\pm0.02$&$0.07\pm0.02$&$0.07\pm0.02$&$0.04\pm0.02$\\
\lbrack Ar III\rbrack&7135.79&1.39&$3.12\pm0.38$&$3.56\pm0.16$&$2.51\pm0.09$&$2.45\pm0.04$&$1.95\pm0.07$&$1.83\pm0.05$\\
\lbrack Fe II\rbrack&7155.16&1.39&$7.44\pm0.36$&$8.30\pm0.13$&$10.04\pm0.11$&$10.97\pm0.08$&$11.88\pm0.11$&$13.26\pm0.12$\\
\lbrack Fe XI\rbrack&7891.94&1.32&$1.88\pm0.35$&$1.85\pm0.16$&$1.94\pm0.09$&$1.87\pm0.07$&$1.89\pm0.07$&$1.71\pm0.09$\\
\lbrack S III\rbrack&9068.6&1.24&-&$2.88\pm0.16$&-&-&-&-\\
\lbrack S III\rbrack&9531.10&1.22&$7.94\pm0.37$&$10.94\pm0.28$&-&-&-&-\\
\hline
Flux(H$\beta$)${}^b$&4861.32&1.71&$48.54\pm0.61$&$164.13\pm1.06$&$468.87\pm2.76$&$515.12\pm1.75$&$556.80\pm2.04$&$564.86\pm1.82$ \\
\hline
\label{tab:broadlines_n}
\end{tabular}
\begin{list}{}{}
 \item[$^{\mathrm{a}}$] Fluxes are relative to the shocked component of H$\beta$ $\times100$
 \item[$^{\mathrm{b}}$] H$\beta$ fluxes are in $10^{-16}\ergsm$
 \item[$^{\mathrm{c}}$] Applies to both lines at $5197.90\ang$ and $5200.26\ang$.
\end{list}
\end{table*}

\begin{table*}
\caption{Fluxes of emission lines from the shocked gas from the southern ER$^{\mathrm{a}}$.}
\begin{tabular}{llccccccc}
\hline
\hline
Ion&$\lambda_{rest}$&Extinct.&5039 d&5703 d&6618 d&6826 d&7183 d&7559 d\\
\hline
\lbrack Ne V\rbrack&3425.86&2.06&-&$4.35\pm1.13$&$2.41\pm0.48$&$3.32\pm0.65$&$1.55\pm0.41$&$1.35\pm0.23$\\
\lbrack N I\rbrack&3466.50&2.05&-&$4.54\pm1.24$&$4.70\pm0.55$&$4.68\pm1.19$&$4.42\pm0.55$&$5.71\pm0.33$\\
\lbrack Ne III\rbrack&3868.75&1.94&$34.39\pm6.30$&$19.26\pm1.00$&$19.76\pm0.80$&$18.20\pm0.50$&$15.93\pm0.50$&$13.38\pm0.48$\\
\lbrack S II\rbrack&4068.60&1.90&$19.59\pm3.78$&$27.43\pm1.37$&$34.39\pm0.93$&$32.24\pm0.52$&$32.05\pm0.45$&$31.14\pm0.57$\\
\lbrack S II\rbrack&4076.35&1.90&$16.24\pm3.78$&$8.90\pm0.60$&$7.75\pm0.43$&$7.76\pm0.22$&$7.75\pm0.20$&$7.46\pm0.26$\\
H$\delta$&4101.73&1.89&$8.38\pm3.53$&$17.64\pm1.18$&$20.94\pm0.61$&$19.02\pm0.30$&$19.45\pm0.33$&$18.72\pm0.40$\\
H$\gamma$&4340.46&1.84&$32.56\pm3.22$&$37.77\pm1.28$&$39.88\pm0.69$&$39.88\pm0.52$&$39.73\pm0.47$&$37.93\pm0.44$\\
\lbrack O III\rbrack&4363.21&1.84&$11.94\pm4.55$&$6.69\pm0.42$&$4.65\pm0.33$&$4.07\pm0.16$&$3.67\pm0.14$&$2.61\pm0.16$\\
\lbrack Fe III\rbrack&4658.05&1.76&-&$2.09\pm0.35$&$1.60\pm0.12$&$1.51\pm0.07$&$1.77\pm0.13$&$1.30\pm0.15$\\
He II&4685.7&1.75&-&$6.89\pm0.80$&$6.71\pm0.34$&$7.29\pm0.17$&$6.91\pm0.14$&$6.16\pm0.15$\\
\lbrack O III\rbrack&4958.91&1.68&$8.34\pm1.69$&$12.22\pm0.47$&$7.74\pm0.20$&$7.14\pm0.32$&$5.74\pm0.23$&$4.86\pm0.12$\\
\lbrack O III\rbrack&5006.84&1.67&$31.42\pm1.99$&$36.96\pm0.96$&$24.28\pm0.26$&$24.04\pm0.31$&$18.89\pm0.20$&$17.38\pm0.20$\\
\lbrack N I\rbrack&5199$^{\mathrm{c}}$&1.63&-&$1.52\pm0.30$&$0.67\pm0.13$&$0.66\pm0.09$&$0.92\pm0.09$&$1.06\pm0.07$\\
\lbrack Fe XIV\rbrack&5302.86&1.61&-&$1.57\pm0.26$&$2.19\pm0.14$&$2.69\pm0.21$&$2.78\pm0.09$&$3.62\pm0.07$\\
\lbrack O I\rbrack&5577.34&1.57&$0.97\pm0.62$&$2.50\pm0.22$&$2.80\pm0.07$&$2.92\pm0.22$&$3.31\pm0.15$&$4.13\pm0.08$\\
\lbrack N II\rbrack&5754.59&1.54&$29.84\pm1.41$&$42.39\pm0.90$&$37.81\pm0.40$&$41.20\pm0.49$&$37.15\pm0.31$&$42.58\pm0.41$\\
He I&5875.63&1.53&$23.52\pm1.89$&$33.29\pm0.72$&$28.82\pm0.33$&$30.04\pm0.53$&$30.09\pm0.23$&$31.68\pm0.26$\\
\lbrack Fe VII\rbrack&6087.0&1.50&-&$0.36\pm0.13$&$0.48\pm0.06$&$0.41\pm0.08$&$0.43\pm0.05$&$0.28\pm0.03$\\
\lbrack O I\rbrack&6300.30&1.47&$24.19\pm1.36$&$46.41\pm0.91$&$43.67\pm0.48$&$48.39\pm0.82$&$47.57\pm0.60$&$54.83\pm0.63$\\
\lbrack S III\rbrack&6312.06&1.47&-&$1.42\pm0.22$&-&-&-&-\\
\lbrack O I\rbrack&6363.78&1.47&$6.70\pm0.66$&$15.28\pm0.59$&$14.70\pm0.53$&$16.53\pm0.43$&$16.09\pm0.17$&$18.99\pm0.30$\\
\lbrack Fe X\rbrack&6374.51&1.47&-&$1.33\pm0.30$&$1.76\pm0.10$&$2.21\pm0.19$&$1.75\pm0.06$&$2.01\pm0.11$\\
\lbrack N II\rbrack&6548.05&1.45&$10.80\pm0.79$&$16.62\pm0.51$&$12.92\pm0.23$&$13.31\pm0.38$&$12.25\pm0.16$&$12.56\pm0.13$\\
H$\alpha$&6562.80&1.45&$241.28\pm7.63$&$388.67\pm5.30$&$364.23\pm2.92$&$384.81\pm2.62$&$366.09\pm1.92$&$411.02\pm2.41$\\
\lbrack N II\rbrack&6583.45&1.45&$35.44\pm1.88$&$50.54\pm0.87$&$39.52\pm0.36$&$42.10\pm0.53$&$36.13\pm0.26$&$37.50\pm0.29$\\
He I&6678.15&1.44&$2.37\pm0.85$&$6.58\pm0.32$&$6.61\pm0.17$&$7.29\pm0.26$&$7.18\pm0.10$&$8.70\pm0.23$\\
\lbrack S II\rbrack&6716.44&1.43&$1.49\pm0.91$&$1.06\pm0.15$&$0.52\pm0.09$&$0.47\pm0.06$&$0.85\pm0.03$&$0.74\pm0.03$\\
\lbrack S II\rbrack&6730.82&1.43&$3.06\pm1.21$&$2.04\pm0.22$&$1.64\pm0.13$&$1.65\pm0.18$&$1.62\pm0.04$&$1.89\pm0.08$\\
\lbrack Ar V\rbrack&7005.67&1.41&-&-&-&-&$0.11\pm0.02$&$0.10\pm0.02$\\
\lbrack Ar III\rbrack&7135.79&1.39&$3.05\pm0.65$&$4.07\pm0.38$&$3.23\pm0.16$&$3.02\pm0.10$&$2.67\pm0.08$&$2.24\pm0.06$\\
\lbrack Fe II\rbrack&7155.16&1.39&$5.12\pm0.59$&$7.38\pm0.34$&$8.19\pm0.13$&$8.92\pm0.10$&$9.59\pm0.08$&$9.94\pm0.12$\\
\lbrack Fe XI\rbrack&7891.94&1.32&-&$1.23\pm0.21$&$1.77\pm0.07$&$1.82\pm0.05$&$1.76\pm0.07$&$1.77\pm0.11$\\
\lbrack S III\rbrack&9068.6&1.24&-&$4.83\pm0.52$&-&-&-&-\\
\lbrack S III\rbrack&9531.10&1.22&$4.81\pm0.84$&$11.00\pm0.83$&-&-&-&-\\
\hline
Flux(H$\beta$)${}^b$&4861.32&1.71&$13.94\pm0.41$&$46.82\pm0.60$&$247.18\pm1.56$&$329.01\pm1.93$&$429.26\pm1.75$&$505.99\pm2.69$\\
\hline
\label{tab:broadlines_s}
\end{tabular}
\begin{list}{}{}
 \item[$^{\mathrm{a}}$] Fluxes are relative to the shocked component of H$\beta$ $\times100$
 \item[$^{\mathrm{b}}$] H$\beta$ fluxes are in $10^{-16}\ergsm$
 \item[$^{\mathrm{c}}$] Applies to both lines at $5197.90\ang$ and $5200.26\ang$.
\end{list}
\end{table*}

\end{document}